\documentclass[11pt,a4wide]{article}

\usepackage{a4wide}

\usepackage{amsmath,amssymb,amsthm}
\usepackage{latexsym,url}

\usepackage{graphicx}

\usepackage{pgf,pgfarrows,pgfnodes}
\usepackage{tikz}
\usetikzlibrary{arrows,automata}
\usepackage{diagbox,xspace}
\usepackage{algorithm,algorithmicx,algpseudocode}

\usepackage{latexsym,amsmath,amssymb,amsfonts}
\usepackage{url}

\usepackage{color}

\usepackage{multicol}

\usepackage{hyperref}

\newtheorem{theorem}{Theorem}[section]

\newtheorem{proposition}[theorem]{Proposition}

\newtheorem{definition}[theorem]{Definition}

\newtheorem{example}[theorem]{Example}

\newcommand{\M}{\mathcal{M}}

\newcommand{\ds}[1]{[\![\mathrm{#1}]\!]}

\newcommand{\agi}{\ensuremath{\mathsf{i}\xspace}}
\newcommand{\aga}{\ensuremath{\mathsf{a}\xspace}}
\newcommand{\agb}{\ensuremath{\mathsf{b}\xspace}}
\newcommand{\agc}{\ensuremath{\mathsf{c}\xspace}}

\newcommand{\coal}[1]{\langle\!\langle{\mathrm{#1}}\rangle\!\rangle}

\newcommand{\coop}[2][]{\langle\!\langle{#2}\rangle\!\rangle_{_{\!\mathit{#1}}}}

\newcommand{\clcoop}[1]{\left[ {#1} \right]\xspace}
\renewcommand{\coop}[2][]{[{#2}]_{_{\!\mathit{#1}}}}

\newcommand{\atlx}{\mathord \mathsf{X}\, }

\newcommand{\atlg}{\mathord \mathsf{G}\, }
\newcommand{\atlu}{\, \mathsf{U} \, }

\newcommand{\ifff}{\leftrightarrow}

\DeclareMathOperator{\Agt}{\mathbb{A}gt} 
\DeclareMathOperator{\Prop}{\Pi} 
\DeclareMathOperator{\Act}{Act} 

\newcommand{\Logicname}[1]{\ensuremath{\mathsf{#1}}}
\newcommand{\CL}{\Logicname{CL}\xspace}

\newcommand{\ATL}{\Logicname{ATL}\xspace}
\newcommand{\ATLs}{\Logicname{ATL^*}\xspace}

\newcommand{\CSR}{\Logicname{ConStR}\xspace}

\newcommand{\defstyle}{\textbf}

\newcommand{\Ax}{\ensuremath{\mathsf{Ax}\xspace}}

\newcommand{\coalCL}[1]{[\mathrm{#1}]}

\newcommand{\condb}[3]{[\mathrm{#1}]_\mathsf{c}(#2 | #3)}
\newcommand{\condd}[3]{\langle \mathrm{#1}\rangle_\mathsf{c}(#2 | #3)}

\newcommand{\condABd}[4]{\langle \mathrm{#1}\rangle_\mathsf{c}(#3; \langle \mathrm{#2} \rangle #4)}

\newcommand{\dere}[4]
{[\mathrm{#1}]_\beta (#3; \langle \mathrm{#2} \rangle  #4)}

\newcommand{\dedic}[4]
{[\mathrm{#1}]_\alpha (#3; \langle \mathrm{#2} \rangle  #4)}

\newcommand{\strat}{\ensuremath{\sigma\xspace}}

\newcommand{\powerset}[1]{\ensuremath{\mathcal{P}({#1})\xspace}}

\newcommand{\fob}{\psi}

\newcommand{\cgm}{\ensuremath{\mathcal{M}\xspace}}

\newcommand{\gmod}{\ensuremath{\cgm}}
\newcommand{\Out}{\ensuremath{\mathsf{Out}\xspace}}

\newcommand{\cgoal}[1]{\langle\![ #1]\!\rangle}

\newcommand{\SFCL}{\Logicname{SFCL}\xspace}

\newcommand{\GPCL}{\Logicname{GPCL}\xspace}

\newcommand{\ocsrAB}{\ensuremath{\mathsf{O}_{c}\xspace}}

\newcommand{\ocsrdr}{\ensuremath{\mathsf{O}_{\beta}\xspace}}
\newcommand{\ocsrdd}{\ensuremath{\mathsf{O}_{\alpha}\xspace}}

\newcommand{\coA}{\ensuremath{\mathrm{A}\xspace}}
\newcommand{\coB}{\ensuremath{\mathrm{B}\xspace}}
\newcommand{\coC}{\ensuremath{\mathrm{C}\xspace}}

\newcommand{\extn}[1]{\| #1\|}
\newcommand{\bisim}{\ensuremath{\mathcal{R}}}

\begin{document}
\sloppypar

\title{A Logic for Conditional Local Strategic Reasoning
\thanks{This paper is a revised and extended version of \cite{DBLP:conf/lori/GorankoJ19}.}
}

\author{Valentin Goranko \\  
Department of Philosophy, Stockholm University, Sweden \\ 
\texttt{valentin.goranko@philosophy.su.se}\\
 \and Fengkui Ju \\
 Faculty of Philosophy, The John Paul II Catholic University of Lublin, 
 Poland \\ 
\texttt{fengkui.ju@kul.pl}}


\maketitle

\begin{abstract}
We consider systems of rational agents who act and interact in pursuit of their individual and collective objectives. 
We study and formalise the reasoning of an agent, or of an external observer, about the 
expected choices of action of the other agents based on their objectives, in order to assess the reasoner's ability, or expectation,  to achieve their own objective.

To formalize such reasoning we extend Pauly's Coalition Logic with three new modal operators of conditional strategic reasoning, thus introducing the Logic for Local Conditional Strategic Reasoning \CSR. We provide formal semantics for the new conditional strategic operators in concurrent game models, introduce the matching notion of bisimulation for each of them, prove bisimulation invariance and Hennessy-Milner property for each of them, and discuss and compare briefly their expressiveness. Finally, we also propose systems of axioms for each of the basic operators of \CSR and for the full logic. 

\end{abstract}

\section{Introduction}
\label{sec:intro}

Consider the following scenario. Alice and Bob are students at DownTown University. Alice is coming to campus today, and has some agenda to complete. Bob wants to meet Alice somewhere on campus today. She does not know that (maybe, even does not know Bob) and they have no communication. Bob may, or may not, know what Alice is going to do on campus, or where and at what time she will go during the day.
Using his knowledge of what, where, and when Alice intends to do today, Bob wants to come up with a plan of how (where and when) to meet her.

From a more general perspective, we consider a  scenario of agents acting independently, and possibly concurrently, in pursuit of their individual and collective goals and we analyse the reasoning of an agent 
(or, observer) 
 about the possible \emph{local} actions (at the current state only) of the other agents and their effect for realising or enabling the outcome of interest for the reasoner.

\paragraph{Related work and motivation.} The kind of strategic reasoning discussed here is within the conceptual thrust motivating the research on logic-based strategic reasoning over the past two decades, starting with Coalition Logic CL (\cite{Pauly01phd}, \cite{Pauly02modal}), its temporal extension, the alternating-time temporal logic ATL (\cite{AHK-02}), its epistemic extension ATEL (\cite{vdHW04}),  and gradually evolving towards increasingly expressive formalisms, such Strategy Logic SL
\cite{tocl/MogaveroMPV14}
(cf. \cite{corr/MogaveroMPV16}).
See \cite{DBLP:series/lncs/8972} and \cite{AgotnesGorankoJamrogaWooldridgeHEL15} for overviews of the area.
{
Most of these logical systems (except SL where the agents' strategies are explicitly named in the language) assume arbitrary or adversarial behaviour of the agents outside of the proponent coalitions in CL, ATL, and ATEL. Also, the knowledge of the agents involved in ATEL refers to truths (of formulae in the language) at the current state, rather than to their knowledge about each others'  objectives and available actions. Thus, these logics formalise \emph{absolute / unconstrained} strategic reasoning -- usually, by an external observer -- about the unconditional strategic abilities of agents and coalitions to achieve their goals. 
However, such unconstrained strategic reasoning is seldom applicable in practice, except in purely adversarial zero-sum games. 
 Usually, all agents acting in the system (except for the environment, or an absolute adversary) have their own goals and act in pursuit of their fulfilment, rather than just to prevent the proponents from achieving their goals. This calls for a more refined strategic reasoning, \emph{conditional}  on the agents' knowledge of the opponents' goals and possible available actions to achieve them, which is the proposal of the present paper. 
It should be noted that there is a recent line of research on \emph{rational synthesis} \cite{tacas/FismanKL10}, \cite{KupfermanPV16} and \emph{rational verification} \cite{aaai/WooldridgeGHMPT16}, \cite{DBLP:journals/ai/GutierrezHW17}, which does take into account all agents' goals, but aims at designing stable strategy profiles (Nash equilibria) that only guarantee the satisfaction of the goal of one special agent (the proponent, representing the system), whereas all others are supposed to act rationally and accept the proposed solution, whether it satisfies their own goals or not. Thus, our work takes an essentially different perspective and has quite different objectives. We are aware of few other works that deal more directly and explicitly with \emph{conditional} strategic reasoning in a sense akin to the present paper. Besides the earlier, conference version \cite{DBLP:conf/lori/GorankoJ19} of this work, perhaps the closest to it in spirit is the recent \cite{GorankoEnqvist18}, to which the present work relates both conceptually and technically, as well as the conceptually related work \cite{NaumovYuan2019}, 
which presents a logic which can express statement of the type: ``The coalition $\mathrm{B}$ has a strategy to achieve their goal once they know the strategy of the coalition $\mathrm{A}$, no matter what the strategy is''. If the epistemic ingredient in it is considered implicit, this statement is expressible by our  operator for `reactive strategic ability' $\ocsrdr$. 
We also note that an axiomatic system is proposed and proved complete in \cite{NaumovYuan2019}, which shares some basic axioms with our axiom system for $\ocsrdr$ presented in Section \ref{subsec:AxiomsOdr}, but differs from it on others, 

\paragraph{Our contributions.}
In this work we identify several patterns of conditional strategic reasoning of an observer or an active agent, depending on his/her knowledge about the objectives and possible actions of the other agents. 
To formalize such reasoning we extend Coalition Logic (\cite{Pauly01phd}, \cite{Pauly02modal}) with three new modal operators of conditional strategic reasoning, thus introducing the Logic for Local Conditional Strategic Reasoning \CSR. 
We provide formal semantics for the new conditional strategic operators, introduce the matching notion of bisimulation for each of them and discuss and compare briefly their expressiveness. We then also propose systems of axioms for each of the basic operators of \CSR and for the whole logic, without stating yet completeness claims (for lack of space,  these are left to future work).

\paragraph{Structure of the paper.}
Section \ref{sec:preliminaries} provides some preliminaries on concurrent game models and the coalition logic CL. Then, 
Section \ref{sec:discussion} presents an informal discussion on conditional strategic reasoning, motivating the further technical work.
Section \ref{sec:logics} introduces three modal operators formalising patterns of conditional strategic reasoning and the new logic \CSR as an extension of Coalition Logic with these operators.
Section \ref{sec:bisimulations} introduces the matching notion of bisimulation for that logic  and discuss briefly it expressiveness.
In Section \ref{sec:Axioms} we propose systems of axioms for each of the basic operators of \CSR and for the full logic.
We end with brief concluding remarks in Section \ref{sec:Concluding}.

\section{Preliminaries}
\label{sec:preliminaries}

\paragraph{Multi-agent game models.}
\label{subsec:models}
%
We fix a finite set of \defstyle{agents} $\Agt = \{a_1,...,a_n\}$ and a set of  \defstyle{atomic propositions} $\Prop$. Subsets of $\Agt$ will also be called \defstyle{coalitions}.
\begin{definition} 
\label{def:gamemodel}
A \defstyle{game model}\footnote{These game models are essentially equivalent to concurrent game models used in  \cite{AHK-02}.} for $\Agt$ and $\Prop$ is a tuple
\[
\gmod = (S,\{\Sigma_{a}\}_{a\in\Agt}, g,V)
\]
where
$S$ is a non-empty set of \defstyle{states};
each $\Sigma_{a}$ is a non-empty set of possible \defstyle{actions} of agent $a$;
$V: \Prop \to \powerset{S}$ is a \defstyle{valuation} of the atomic propositions from $\Prop$ in $S$;
and $g$ is a \defstyle{game map} that assigns to each $s \in S$ a strategic game form $g(s) = (\Sigma^{s}_{a_1},....\Sigma^{s}_{a_n},o_{s})$,
where each $\Sigma^{s}_{a_i} \subseteq \Sigma_{a_i}$ is a non-empty set of actions available to player $a_i$ at $s$, and
\[
o_{s} : \Sigma^{s}_{a_1} \times ... \times \Sigma^{s}_{a_n} \to S
\]
is a \defstyle{local outcome function} assigning to any \defstyle{action profile} $\strat \in \Sigma^{s}_{a_1} \times ... \times \Sigma^{s}_{a_n}$ the \defstyle{outcome state} $o_{s}(\strat)$ produced by $\strat$ when applied at $s\in S$. The set $\Sigma^{s}_{a_1} \times ... \times \Sigma^{s}_{a_n}$ of \defstyle{action profiles available at $s$} will be denoted by $\Act_s$.

Now, the \defstyle{global outcome function} in $\gmod$ is the partial mapping
\[
O : S \times \Sigma_{a_1} \times ... \times \Sigma_{a_n} \dashrightarrow S
\]
defined by $O(s,\strat) = o_{s}(\strat)$, whenever $\strat \in \Act_s$.

Given a coalition $\coC \subseteq \Agt$, a \defstyle{joint action} for $\coC$ in the model $\gmod$ is a tuple of individual actions $\strat_{\coC} \in \prod_{a \in \coC} \Sigma_a$. For any such joint action $\strat_{\coC}$ that is available at $s \in S$, we define the \defstyle{set of outcome states from $\strat_{\coC}$ at $s$}:
\[
\Out[s,\strat_{\coC}] = \left\{u \in S \mid \exists \strat \in \Act_s:
\; \strat \vert_{\coC} = \strat_{\coC} \; \& \; o_{s}(\strat) = u \right\}
\]
where $\strat \vert_{\coC}$ is the restriction of $\strat$ to $\coC$. Note that the empty tuple $\sigma_\emptyset$ is the only available joint action for the empty coalition $\emptyset$ at any state.
\end{definition}

\paragraph{The basic logic for coalitional strategic reasoning \CL.}
\label{sec:thelogic}

The Coalition Logic 
\CL was introduced in \cite{Pauly01phd}, cf. also \cite{Pauly02modal}. \CL extends the classical propositional logic  with \textit{coalitional strategic modal operators}  $\clcoop{\coC}$, for any coalition of agents $\coC$. The formulae of \CL are defined as follows:
\[
\varphi := p\ |\ \neg\varphi \ |\ \varphi_1 \lor \varphi_2\ |\ \clcoop{\coC} \varphi
\]
We will write $\clcoop{\agi}$ instead of $\clcoop{\{\agi\}}$. The intuitive reading of {$\clcoop{\coC} \varphi$} is:

\begin{quote}
``{The coalition $\coC$ \emph{has a joint action that ensures an outcome (state) satisfying $\varphi$}, regardless of how all other agents act.}''
\end{quote}

The semantics of \CL is defined in terms 
of the notion of \defstyle{truth of a \CL-formula $\fob$ at a state $s$ of a game model $\cgm$}, denoted ${\cgm,s \vDash \fob}$, by induction on formulae, via the key clause:

\begin{description}
\item[{$\gmod,s\models \clcoop{\coC}\phi ~ \Leftrightarrow$}] there exists a joint action $\sigma_{\coC}$ available at $s$, such that $\gmod,u\models \phi$ for each $u \in \Out[s,\sigma_{\coC}]$
\end{description}
Thus, $\clcoop{\coC}\phi$ formalises a claim of the ability of the agent/coalition \coC\ to choose a suitable (joint) action to ensure achieving the goal $\phi$ \emph{regardless of how all other agents choose to act}, and therefore without assuming that the agents in \coC\ know the goal(s) of the remaining agents and their available actions to achieve these goals. 

The notion of \emph{bisimulation} that guarantees truth invariance of all \CL-formulae was first defined in  \cite{Alur98refinement} for the so called `alternating transition systems' (equivalent to a special case of concurrent game models), then independently in \cite{Pauly01phd} for the abstract game models defined there,  and later in \cite{Agotnes07irrevocable} for concurrent game models, which definition we give here. 

\begin{definition}[\CL-bisimulation]
Let $\gmod = (S,\{\Sigma_{a}\}_{a\in\Agt},g,V)$ be a concurrent game model. A binary relation $\bisim \subseteq S^{2}$ is a \defstyle{\CL-bisimulation in \gmod} if it satisfies the following conditions for every pair of states $(s_1, s_2)$ such that $s_1 \bisim s_2$ and for every coalition $\coC$:

\begin{description}
\itemsep = 1pt

\item[\textbf{Atom equivalence:}] For every $p \in \Prop$:  $s_1 \in V(p)$ iff $s_2 \in V(p)$.

\item[\textbf{Forth:}] For every joint action $\strat^{1}_{\coC}$ of $\coC$ at $s_1$, there is a joint action $\strat^{2}_{\coC}$ of $\coC$ at $s_2$ such that for every $u_{2} \in \Out[s_{2},\strat^{2}_{\coC}]$ there exists $u_{1} \in \Out[s_{1},\strat^{1}_{\coC}]$ such that $u_1 \bisim u_2$.

\item[\textbf{Back:}] Like \textbf{Forth}, but with $1$ and $2$ swapped.
\end{description}
\end{definition}

\noindent \CL-bisimulation is defined here \emph{within} a model. It readily extends to \CL-bisimulation \emph{between} models, by treating both as parts of a single model.

\paragraph{The Alternating-time Temporal Logic $\ATLs$.}  
\label{subsec:ATL}

The Alternating-time Temporal Logic $\ATLs$, proposed by \cite{AHK-02}, is an extension of \CL with temporal operators. Its featured operator is $\coal{\coC} \phi$, denoting the claim that $\coC$ has a joint strategy that guarantees the satisfaction of $\phi$, where  $\phi$ is a `temporal objective', i.e. a (path) formula beginning with a temporal operator `\emph{nexttime}' $\atlx$, `\emph{always}' $\atlg$, or `\emph{until}' $\atlu$. The logic \CL embeds into \ATLs as the fragment extending propositional logic only with combinations of strategic and temporal operators of the type $\coal{\coC} \atlx$, cf \cite{Goranko01}. We only mention \ATLs here for the sake of some further references, but the present paper will not make any essential use of that logic, and no familiarity with it, nor even with its fragment \ATL, is required.

\newpage
\section{Conditional strategic reasoning: an informal discussion} 
\label{sec:discussion}

\footnote{The reader who is only interested in the technical part of the paper can skip this section without essential loss.}Recall the scenario with the students Alice and Bob. Suppose that Alice has an objective $\gamma_{A}$ to achieve  -- say, to meet with her supervisor Carl on campus today.
Suppose also that Alice has several possible choices of an action (or a `strategy')
\footnote{In this paper we focus on local reasoning, about once-off actions, but in this section the word `action' can be conceived in a wider sense, and may mean either a once-off action, or a global strategy guiding the long term behaviour of the agent.}
that would possibly, or certainly, guarantee the achievement of her objective. 
In our example, suppose these choices are: meeting in the supervisor's office, or in the library, or at the campus caf\'e. 
 
\subsection{Observer's conditional reasoning about an outcome from agent's actions}
\label{subsec:observer}

Let us first consider the case where Bob is just an observer who is not acting, but only reasoning about the consequences from Alice's possible actions with respect to the occurrence of another -- intended or not -- outcome event $\gamma_{B}$.
For instance, suppose that Bob is interested in meeting with Alice on campus today 
-- let us call that event $\gamma_{B}$ -- and is sitting in the campus caf\'e and reasoning about whether Alice will happen to come to the caf\'e, thus enabling the event $\gamma_{B}$ (recall that Alice may not know about bob expecting her there, or at all).
More generally, we can also assume that there are other agents, besides Alice, also acting in pursuit of their own goals, and Bob is reasoning about their individual and collective choices of action and the consequences from these choices. This leads to an \emph{observer's conditional strategic reasoning}
about claims of the type:

\smallskip
``\textit{Some/every action of Alice that guarantees achievement of $\gamma_{A}$ also
guarantees/enables occurrence of the outcome $\gamma_{B}$}".

\smallskip
Depending on Bob's knowledge about Alice's objective and of her expected choices of action there can be several possible cases for Bob's reasoning about the expected occurrence of the outcome $\gamma_{B}$.

\subsubsection{Observer Bob's reasoning, case 1: Bob knows nothing about Alice}
\label{subsec:case1}

Suppose that \textit{Bob does not know Alice's objective $\gamma_{A}$}, and therefore has no a priori expectations about her choice of action.
In our example, suppose that Bob only knows that Alice is coming to campus today, but not why and where on campus she is going. 
Then, Bob can \emph{only} claim for sure that the outcome $\gamma_{B}$ will occur if $\gamma_{B}$ is inevitable, regardless of how Alice (and all others) will act.
For instance, if Bob knows that Alice is coming to campus and he is standing by the only entrance of the campus, then he will know for sure that he is going to meet Alice ($\gamma_{B}$ will occur), no matter what she will do there.
This claim can be expressed in Coalition Logic \CL 
simply as $[\emptyset] \gamma_{B}$.

\subsubsection{Observer Bob's reasoning, case 2: Bob only know Alice's goal}
\label{subsec:case2}

Suppose now that Bob \textit{does know Alice's objective} and knows that Alice can guarantee the achievement of that objective and will act towards that, but Bob does not know \emph{how exactly} Alice might act.
{E.g., Bob knows that Alice is coming to campus 
 to meet with her supervisor Carl but does not know where and when.}
Then, Bob can only claim that the outcome $\gamma_{B}$ will occur for sure if $\gamma_{B}$ is true \emph{on every possible course of events (``play")
on which $\gamma_{A}$ is true}.
For instance, if Bob knows that Alice's supervisor will be working in his office for the whole day, and he is sitting in the corridor, next to Carl's office, then he knows that he will meet with Alice ($\gamma_{B}$ will occur) no matter when Alice comes to meet with Carl (i.e. no matter how $\gamma_{A}$ occurs).

This can be expressed in \CL simply as $[ \emptyset ] (\gamma_{A} \to \gamma_{B})$ and 
reflects the case when $\gamma_{A}$ can occur in various, possibly unintended ways,  but its occurrence always implies the occurrence of $\gamma_{B}$ (e.g. if Bob is with Carl throughout the day, then even if Alice bumps into Carl accidentally, Bob will still meet her).

\subsubsection{Observer's reasoning, case 3: Bob knows Alice's goal and actions}
\label{subsec:case3}

Suppose now that Bob not only knows Alice's objective, but also \emph{knows all possible actions (or, strategies)} of Alice that can ensure the satisfaction of her objective $\gamma_{A}$, and  \emph{knows that Alice will perform one of them}, but  \emph{does not know to which one}.
(E.g., Bob knows that Alice, who is coming to campus to meet with her supervisor,  can meet with him either in his office, or in the library, or in the caf\'e.)
Now, for Bob to claim that the outcome $\gamma_{B}$ will occur for sure, it suffices to know that \emph{each action of Alice that guarantees $\gamma_{A}$} will also guarantee $\gamma_{B}$. 
(E.g., suppose that all possible meeting places for Alice and her supervisor are in the main building and Bob is waiting at the only entrance of the main building.)  

Here the conditional ``If $\gamma_{A}$ then $\gamma_{B}$'' has a  suitably constrained context, 
specifying that $\gamma_{A}$ can occur only because the agent (Alice) takes a deliberate  action to bring about $\gamma_{A}$. 
This can no longer be expressed in \CL and requires introducing a new strategic operator.

Lastly, suppose that \textit{Bob also knows the specific action which Alice is taking in order to guarantee the achievement of her goal}.
Then, Bob can claim that the outcome $\gamma_{B}$ will occur for sure, as long as \emph{that specific action} of Alice guarantees the satisfaction of $\gamma_{B}$.
To formalise that one needs explicit names for actions, but in our logic we will be able to state something stronger, viz. that \emph{every specific action of Alice that guarantees $\gamma_{A}$} will also bring about  $\gamma_{B}$.

\subsection{Conditional reasoning of an agent about another agent's actions}
\label{subsec:agent}
 
Suppose now that Bob is not just a passive observer, but an acting agent, who has the outcome $\gamma_{B}$ as his own goal. There may be other agents, besides Alice and Bob, also acting in pursuit of their own goals, and Bob is reasoning about their expected 
choices of action and the consequences from these choices.
Now, Bob is to decide -- based on his reasoning about Alice's (and other agents') possible choices of actions --  on his own action in pursuit of $\gamma_{B}$.
This calls for an  \emph{agent's conditional strategic reasoning} about statements of the type:

\smallskip
``\textit{For some/every action of Alice that guarantees achievement of $\gamma_{A}$, Bob has
an action of his own to guarantee achievement of his objective $\gamma_{B}$}".
\smallskip

We call this \emph{local conditional strategic reasoning}, as it only refers to the immediate actions of the agents, not about their \emph{global strategies}.
Respectively, the outcomes from the local action profiles are just successor states, while in the general case they are (finite or possibly infinite) \emph{plays}. The global conditional strategic reasoning will be treated in a follow-up work.

\medskip
Each of the cases considered in Section \ref{subsec:observer} accordingly applies here, too, with the only difference being that now Bob is to choose a suitable action of his own. Besides, there are several additional cases to consider regarding the possible choice of action of Bob.

\subsubsection{Agent Bob's reasoning, case 1: \emph{assuming Alice's cooperation}.}
\label{subsec:coop}

First, suppose, in addition, that Alice also knows Bob's objective and can choose to cooperate with Bob by selecting a suitable action $\strat_\coA$ that would not only guarantee achievement of her objective but will also enable  Bob to supplement $\strat_\coA$ with an action $\strat_\coB$ which would then also guarantee achievement of his objective, too. (So, we also assume that Alice knows enough about Bob's possible actions.)
We refer the reader to Example \ref{example1} for a formal model illustrating the agent's ability assuming cooperation from the other agent.

\medskip

\subsubsection{Agent Bob's reasoning \emph{not assuming Alice's cooperation}.}
\label{subsec:nocoop}

Now, suppose Bob cannot count on Alice's cooperation.
Still, the statement
\begin{quote}
``\textit{Whichever way Alice acts towards achieving the objective $\gamma_{A}$, Bob can act so as to bring about achievement of his objective $\gamma_{B}$.}''
\end{quote}
admits two different readings, which we respectively call 
`\emph{proactive ability}'  and `\emph{reactive ability}' 
which we discuss below\footnote{In \cite{DBLP:conf/lori/GorankoJ19} these were called respectively `\emph{ability de re}'  and `\emph{ability de dicto}'}.

\subsubsection{Agent Bob's reasoning, case 2: reactive ability}
\label{subsec:dere}
 
In this case, \emph{for every action} of Alice that ensures $\gamma_{A}$ Bob is to choose \emph{reactively} an action of his, \emph{generally dependent on Alice's action}, that would also ensure the occurrence of $\gamma_{B}$ (possibly in different ways for the different actions).
This essentially corresponds to the case where Bob knows Alice's action at the time of choosing his own action, and for every choice of action of Alice that guarantees $\gamma_{A}$, Bob's respective choice will also bring about  $\gamma_{B}$. 
For instance, in our running example, if Bob knows where Alice is going to meet with 
 her supervisor, he can choose respectively where to go and wait for her. 

More formally, each of Alice's actions that would guarantee $\gamma_{A}$ generates a set of \emph{possible} outcome states (plays), and for each such set Bob is looking for a respective action that will bring about $\gamma_{B}$ on that set of outcome states.

\subsubsection{Agent Bob's reasoning, case 3: proactive ability}
\label{subsec:dedic}

The case of \emph{proactive ability} is when Bob only knows that Alice has committed to act so as to achieve her goal $\gamma_{A}$ (to meet with Carl at one of the 3 possible meeting places), but does not knows the action that Alice has chosen and her choice will remain unknown to Bob at the time when he is to choose his action aiming at satisfying $\gamma_{B}$ (meet Alice).  

In this case Bob must consider all possible courses of events (plays) that can occur as a result of Alice acting towards achieving $\gamma_{A}$ and reason about whether
he can choose \emph{proactively and uniformly} one action that would bring about $\gamma_{B}$ regardless of which action  Alice may choose to apply in order to achieve her goal $\gamma_{A}$.   
For instance, in our running story, assuming that all meeting places are in the main building,  
Bob can choose to wait for Alice at the only entrance of that building.
Formally speaking, in this case, based on his knowledge Bob considers
the set of states in the model which is the union of all sets of outcome states enabled by the specific actions of Alice that would guarantee $\gamma_{A}$, and is looking for an action that will bring about $\gamma_{B}$ on each of these outcome states.

\smallskip 
We refer the reader to Example \ref{example2} for a formal model illustrating the concepts of proactive and reactive abilities of an agent and their difference.

\smallskip 
Top sum up, the  \emph{proactive -- reactive} ability distinction applies to Bob depending on whether or not he knows Alice's choice at the time when he is to make his own choice of action. If he knows Alice's choice at that time, his reasoning is about reactive ability,  
else it is a proactive ability reasoning.
We note that the notions of proactive and reactive ability respectively correspond to the notions of \emph{$\alpha$-effectivity} and \emph{$\beta$-effectivity} in game theory (cf. e.g. \cite{Abdou91effectivity}).  

\smallskip 
Lastly, an important point: even though knowledge of the agent about the other's goals and possible actions is essential, it will not feature in our formal logical language, nor in the formal semantics, but only in the external reasoners' analysis of which case of conditional strategic ability applies.

\section{The logic of conditional strategic reasoning \CSR}
\label{sec:logics}

\subsection{Modal operators for conditional strategic reasoning}

Given coalitions $\coA$ and $\coB$ and joint actions $\strat_\coA$ for $\coA$ and $\strat_\coB$ for $\coB$, we say that \defstyle{$\strat_\coB$ is consistent with $\strat_\coA$} if $\strat_\coB$ coincides with $\strat_\coA$ on $\coA \cap \coB$.

We now introduce new operators for conditional strategic reasoning, for any coalitions $\coA$ and $\coB$, with intuitive semantics corresponding to the three reasoning cases in Section \ref{subsec:agent}, as follows.

\medskip
(\ocsrAB) \
$\condABd{\coA}{\coB}{\phi}{\psi}$ says that $\coA$ has a joint action $\strat_\coA$ which, when applied, guarantees the truth of $\phi$ and enables $\coB$ to apply a joint action $\strat_\coB$ that is consistent with $\strat_\coA$ and guarantees $\psi$ when \emph{additionally} applied by $\coB$, in sense that all agents in $\coA$ act according to $\strat_\coA$ and those in $\coB \setminus \coA$ act according to $\strat_\coB$\footnote{We note that $\condABd{\coA}{\coB}{\phi}{\psi}$ is equivalent to $\coal{\coA}(\phi \land \coal{B \setminus \coA}\psi)$ in \ATLs.} 
This operator formalises the agent's reasoning Case 1 discussed in Section \ref{subsec:agent}, where $\coA$ knows the objective of $\coB$ and can choose to cooperate with $\coB$ by selecting a suitable action.

\medskip
(\ocsrdd) \
$\dedic{\coA}{\coB}{\phi}{\psi}$ says that the coalition $\coB \setminus \coA$ has an action $\strat_{\coB \setminus \coA}$ such that if $\coA$ applies any action that guarantees the truth of $\phi$, then $\coB \setminus \coA$ can guarantee the truth of $\psi$ by applying additionally the action $\strat_{\coB \setminus \coA}$.

This operator formalises a claim of the ability of the agent/coalition \coB\ to choose a suitable  (joint)  action so as to achieve the goal $\psi$ assuming that \coA\ acts so as to achieve the goal $\phi$, if \coB\ is to choose their  (joint)  action \emph{before} \coA\ chooses their  (joint) action, or before \coB\ learns the action of \coA. This corresponds to the notion of agent's \emph{proactive ability} discussed in Section \ref{subsec:dedic}, respectively to the game-theoretic notion of \emph{$\alpha$-effectivity}, hence the notation.

\medskip
(\ocsrdr) \
$\dere{\coA}{\coB}{\phi}{\psi}$ says that for any joint action  $\strat_\coA$ of $\coA$ that guarantees the truth of $\phi$, when applied by $\coA$ there is an action $\strat_\coB$ that is consistent with $\strat_\coA$ and guarantees $\psi$ when additionally applied by $\coB$\footnote{
Note that $\dere{\coA}{\coB}{\bot}{\psi}$ is vacuously true for any \coA, \coB, and $\psi$, as then there cannot be such joint actions $\strat_\coA$ that enable satisfying $\bot$. This may sounds odd, but it is no special phenomenon in \CSR, as the same effect occurs  in FOL with universal quantification over an empty set of objects.}. This operator formalises a claim of the ability of the agent/coalition \coB\ to choose a suitable  (joint) action so as to achieve the goal $\psi$ assuming that \coA\ acts so as to achieve the goal $\phi$, if \coB\ is to choose their  (joint) action \emph{after} \coB\ learns the (joint) action of \coA. This corresponds to the notion of agent's \emph{reactive ability} discussed in Section \ref{subsec:dere}, respectively to the game-theoretic notion of \emph{$\beta$-effectivity}, hence the notation.

\subsection{The language of \CSR}

We fix a finite nonempty set of agents $\Agt$  and a countable set of atomic propositions $\Prop$. The formulae of $\CSR$, where $p \in \Prop$ and $\coA, \coB \subseteq \Agt$ are defined as follows:
\[
\phi ::=
p \mid
\top \mid
\neg \phi \mid
(\phi \land \phi) \mid
\condABd{\coA}{\coB}{\phi}{\phi}
\mid 
\dedic{\coA}{\coB}{\phi}{\phi}
\mid 
\dere{\coA}{\coB}{\phi}{\phi}
\]

\subsection{Some definable operators and expressions in \CSR}

The following can be easily seen from the informal semantics above, and can also be easily verified with the formal semantics introduced further.

\begin{itemize}

\item The coalitional operator from \CL is definable as a special case of each of \ocsrAB,  \ocsrdd, \ocsrdr\ as follows: 

\begin{itemize}
\item[(\ocsrAB)] 
$\coalCL{\coA} \phi := \condABd{\coA}{\coA}{\phi}{\phi}$ or
$\coalCL{\coA} \phi := \condABd{\coA}{\coA}{\phi}{\top}$;

\item[(\ocsrdd)]
$\coalCL{\coA} \phi :=  \dedic{\emptyset}{\coA}{\top}{\phi}$. This expresses the case when \coB\ is not informed about the goal of \coA\ and has to choose proactively a joint action, before \coA\ has chosen their action. 
Thus, it indeed claims an unconditional ability of \coB\ to choose an action that guarantees $\phi$. 

\item[(\ocsrdr)] 
$\coalCL{\coA} \phi :=  \dere{\emptyset}{\coA}{\top}{\phi}$, or
$\coalCL{\coA} \phi :=  \dere{\overline{\coA}}{\coA}{\top}{\phi}$;
\ 
(the empty coalition has only one strategy, and it guarantees the satisfaction of $\top$)
\end{itemize}

Thus, the cases of observer's reasoning discussed in Sections 
\ref{subsec:case1} and \ref{subsec:case2} are readily formalisable in \CSR.

\smallskip
\item The dual to \ocsrAB\ operator 
$\lnot \condABd{\coA}{\coB}{\phi}{\lnot \psi}$
says that every joint action of $\coA$ that, when applied, guarantees the truth of $\phi$, would prevent $\coB$ from acting additionally so as to guarantee $\psi$.
This formalises the conditional reasoning scenario where the goals of $\coA$ and $\coB$ are conflicting and where Bob can establish that whichever way $\coA$ acts towards their goal, that would block $\coB$ from acting to guarantee achievement of its goal.

\smallskip
\item $\dere{\coA}{\coB}{\top}{\psi}$ essentially formalises the case when \coB\ is not informed about the goal of \coA,  but has to choose their action after learning the  action of \coA.

\smallskip
\item
On the other hand, 
$\condb{\coA}{\phi}{\psi} := \dere{\coA}{\emptyset}{\phi}{\psi}$, also equivalent to  
$\dedic{\coA}{\emptyset}{\phi}{\psi}$, says that for any joint strategy of $\coA$, if it guarantees $\phi$ to be true, then it guarantees $\psi$ to be true, too. That 
formalises the case in section \ref{subsec:case3} of reasoning of an observer who knows both the goal $\phi$ and the possible actions of \coA,  about the occurrence of outcome $\psi$.

\smallskip

\item $\condd{\coA}{\phi}{\psi} := \lnot \condb{\coA}{\phi}{\lnot \psi}$ says that there is a joint strategy of $\coA$ that guarantees $\phi$ to be true and enables $\psi$ to be true, too. Note that it is equivalent to a special case of the ``socially friendly coalitional operator" \textsf{SF}, $\coop{\coC}(\phi; \psi_{1},\ldots,\psi_{k})$, introduced in \cite{GorankoEnqvist18}, viz. $\condd{\coA}{\phi}{\psi} \equiv \coop{\coA}(\phi; \psi)$.

\smallskip

Moreover, $\condd{\coA}{\phi}{\psi}$ is also definable as $\condABd{\coA}{\overline{\coA}}{\phi}{\psi}$, where $\overline{\coA} = \Agt \setminus \coA$.

\smallskip

\item The coalitional operator $[\coA]$ from \CL is a special case of the above:
$\coalCL{\coA} \phi :=\condd{\coA}{\phi}{\top}$, meaning ``$\coA$ has a joint action to ensure the truth of $\phi$"\footnote{NB: We have preserved the box-like notation for $[\coA]$ from \CL, even though it is not consistent with ours.}.

\smallskip

\item $\condABd{\coA}{\coB}{\phi}{\psi}$ is definable in terms of the ``group protecting coalitional operator" \textsf{GIP}, introduced in \cite{GorankoEnqvist18}:
$\condABd{\coA}{\coB}{\phi}{\psi} \equiv \cgoal{A \triangleright \phi, \coA \cup \coB \triangleright \psi}$.
Nevertheless, it now has a different motivation and intuitive interpretation.
\end{itemize}

\subsection{Formal semantics of $\CSR$}

Given coalitions $\coA, \coB \subseteq \Agt$ and joint actions $\strat_\coA$ for $\coA$ and $\strat_\coB$ for $\coB$, we define $\strat_\coA \uplus \strat_\coB$ to be the joint action for $\coA \cup \coB$ which equals to $\strat_\coA$ when restricted to $\coA$ and equals to $\strat_\coB \vert_{\coB \setminus \coA}$ when restricted to $\coB \setminus \coA$. 
Thus, in particular, $\strat_\coA \uplus \strat_\coB =\strat_\coA$ for any 
$\coB \subseteq \coA \subseteq \Agt$.

Now, let $\gmod = (S,\{\Sigma_{a}\}_{a\in\Agt}, g,V)$ be a game model. The formal semantics of \CSR extends the one of \CL to the new operators as follows:

\begin{description}
\itemsep = 3pt

\item[{$\cgm, s \Vdash \condABd{\coA}{\coB}{\phi}{\psi} ~ \Leftrightarrow$}]
$\coA$ has a joint action $\strat_\coA$, such that \\
$\M, u \Vdash \phi$ for every $u \in \Out[s,\strat_\coA]$ and $\coB$ has a joint action $\strat_\coB$ \\
such that $\M, u \Vdash \psi$ for every $u \in \Out[s,\strat_\coA \uplus \strat_\coB]$.

\item[{$\cgm, s \Vdash \dedic{\coA}{\coB}{\phi}{\psi} ~ \Leftrightarrow$}]
$\coB$ has a joint action $\strat_\coB$ such that \\
for every joint action $\strat_\coA$ of $\coA$, if $\M, u \Vdash \phi$ for every $u \in \Out[s,\strat_\coA]$, \\
then $\M, u \Vdash \psi$ for every $u \in \Out[s,\strat_\coA\uplus \strat_\coB]$.

\item[{$\cgm, s \Vdash \dere{\coA}{\coB}{\phi}{\psi} ~ \Leftrightarrow$}]
for every joint action $\strat_\coA$ of $\coA$ such that $\M, u \Vdash \phi$ for every $u \in \Out[s,\strat_\coA]$,
$\coB$ has a joint action $\strat_\coB$ (generally, dependent on $\strat_\coA$) such that $\M, u \Vdash \psi$ for every $u \in \Out[s,\strat_\coA \uplus \strat_\coB]$.
\end{description}

\textit{Remark:} The semantics of each of the operators above can be re-stated to consider joint actions for $\coB \setminus \coA$ rather than the whole $\coB$. For instance, it can be easily verified for the latter operator, that $\cgm, s \Vdash \dedic{\coA}{\coB}{\phi}{\psi}$ iff $\coB \setminus \coA$ has a joint action $\strat_{\coB \setminus \coA}$ such that for every joint action $\strat_\coA$ of $\coA$, if $\M, u \Vdash \phi$ for every $u \in \Out[s,\strat_\coA]$, then $\M, u \Vdash \psi$ for every $u \in \Out[s,\strat_\coA\uplus \strat_{\coB \setminus \coA}]$.

\subsection{Some examples}
\label{subsec:ex}

Here we provide a few simple examples illustrating the semantics of $\CSR$.

\begin{example}
\label{example1}
The game model $\M$ below has two players, $\aga$ and $\agb$. Each has two actions at state $s_0$: $a_1, a_2$, resp. $b_1, b_2$.

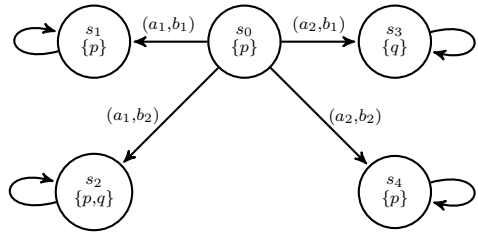
\begin{center}
\begin{tikzpicture}[->,>=stealth',shorten >=1pt,node distance=25mm,thick,every node/.style={transform shape},scale=0.8]
\tikzstyle{every state}=[draw=black,text=black,minimum size=12mm]

\node[state] (0) {$s_0 \atop \{p\}$};
\node[state] (1) [left of=0] {$s_1 \atop \{p\}$};
\node[state] (2) [below of=1] {$s_2 \atop \{p,q\}$};
\node[state] (3) [right of=0] {$s_3 \atop \{q\}$};
\node[state] (4) [below of=3] {$s_4 \atop \{p\}$};

\path 
(1) edge [loop left] node {} (1)
(2) edge [loop left] node {} (2)
(3) edge [loop right] node {} (3)
(4) edge [loop right] node {} (4)

(0) edge [above] node {$_{({a_1}, b_1)}$} (1)
(0) edge [left] node {$_{({a_1}, {b_2})}$} (2)
(0) edge [above] node {$_{({a_2}, b_1)}$} (3)  
(0) edge [right] node {$_{({a_2}, b_2)}$} (4);

\end{tikzpicture}
\end{center}

\noindent It is easy to see that $\M, s_0 \Vdash \condABd{\aga}{\agb}{p}{q}$, while $\M, s_0 \not \Vdash \coalCL{\agb} q$. 
\\
Thus, an agent may have only conditional ability to achieve its goal.
\end{example}

\begin{example}
\label{example2}

The game model $\M$ below has two players, $\aga$ and $\agb$. The agent $\aga$ has 3 actions at state $s_0$, $a_1, a_2, a_3$, and $\agb$ has 2 actions, $b_1, b_2$.

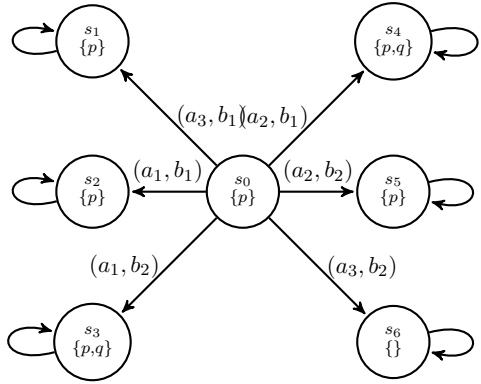
\begin{center}
\begin{tikzpicture}[->,>=stealth',shorten >=1pt,node distance=25mm,thick,every node/.style={transform shape},scale=0.8]
\tikzstyle{every state}=[draw=black,text=black,minimum size=12mm]

\node[state] (0) {$s_0 \atop \{p\}$};
\node[state] (2) [left of=0] {$s_2 \atop \{p\}$};
\node[state] (1) [above of=2] {$s_1 \atop \{p\}$};
\node[state] (3) [below of=2] {$s_3 \atop \{p,q\}$};

\node[state] (5) [right of=0] {$s_5 \atop \{p\}$};
\node[state] (4) [above of=5] {$s_4 \atop \{p,q\}$};
\node[state] (6) [below of=5] {$s_6 \atop \{\}$}; 
  
\path
(1) edge [loop left] node {} (1)
(2) edge [loop left] node {} (2)
(3) edge [loop left] node {} (3)
(4) edge [loop right] node {} (4)
(5) edge [loop right] node {} (5)
(6) edge [loop right] node {} (6)

(0) edge [right] node {$(a_3, b_1)$} (1)
(0) edge [above] node {$(a_1, b_1)$} (2)
(0) edge [left] node {$(a_1, b_2)$} (3)
(0) edge [left] node {$(a_2, b_1)$} (4)
(0) edge [above] node {$(a_2, b_2)$} (5)
(0) edge [right] node {$(a_3, b_2)$} (6);
 
\end{tikzpicture}
\end{center}

Note that: 
\begin{itemize}
\item $\cgm, s_0 \Vdash \dere{\aga}{\agb}{p}{q}$. Indeed, agent $\aga$ has two actions at state $s_0$ to ensure $p$: $a_1$ and $a_2$. For each of them, $\agb$ has an action to ensure $q$, viz.: choose $b_2$ if $\aga$ chooses $a_1$, 
and choose $b_1$ if $\aga$ chooses $a_2$. 

\item  $\cgm, s_0 \not\Vdash \dedic{\aga}{\agb}{p}{q}$. Indeed, neither $b_1$ nor $b_2$ ensures $q$ against both choices $a_1$ and $a_2$ of $\aga$. Thus,  
$\agb$ does not have a uniform action to ensure  $q$ against any action of $\aga$ that ensures $p$. 

 Therefore, $\dedic{\aga}{\agb}{p}{q}$ and $\dere{\aga}{\agb}{p}{q}$ are semantically different. 

\item However, if the outcomes of $({a_2}, {b_1})$ and $({a_2}, {b_2})$ are swapped, then $\dedic{\aga}{\agb}{p}{q}$ becomes true at $s_0$ in the resulting model.

\item 
$(\cgm, s_0)$ does not satisfy the \ATLs formula $\ds{\aga} (\mathrm{X} p \to \coal{\agb} \mathrm{X} q)$ (where $\ds{\coC} \phi := \lnot \coal{\coC} \lnot \phi$), hence it is  not equivalent to  $\dere{\aga}{\agb}{p}{q}$. 
\end{itemize}
\end{example}

\section{Bisimulations and expressiveness of \CSR} 
\label{sec:bisimulations}

\subsection{Bisimulations for \CSR} 
\label{sec:bis-CSR}

The definition of \CSR-bisimulation involves, besides atomic equivalence, 3 nested Forth and Back conditions, for each of the respective new operators \ocsrAB, \ocsrdd, and \ocsrdr\footnote{Each of these conditions is a respective variation of the bisimulation conditions for the basic strategic operators in the logics \SFCL and  \GPCL defined in \cite{GorankoEnqvist18}.}. As the definition of \CL-bisimulation given in Section \ref{sec:preliminaries}, we only define \CSR-bisimulation within a game model, which generalises to \CSR-bisimulation between game models easily. 
Note that the nested back-and-forth conditions are needed because of the patterns of quantification in the semantic definitions of the new strategic operators: first quantification over the actions of \coA\ and \coB, and then over the outcomes generated by these actions.

\begin{definition}[\CSR-bisimulation]
\label{def:CSRbisimulation}
Let $\gmod = (S,\{\Sigma_{a}\}_{a\in\Agt}, g,V)$ be a game model. A binary relation $\bisim \subseteq S^{2}$ is a \defstyle{\CSR-bisimulation in \gmod} if it satisfies the following conditions for every pair of states $(s_1, s_2)$ such that $s_1 \bisim s_2$ and for every coalitions $\coA$ and $\coB$: 

\begin{description}
\itemsep = 1pt

\item[\textbf{Atom equivalence:}] For every $p \in \Prop$:  $s_1 \in V(p)$ iff $s_2 \in V(p)$.

\item[\textbf{\ocsrAB-bisimulation:}] ~ (For illustration, see Figure \ref{fig-bis-c}.)

\begin{figure}[h]
\begin{center}
\begin{tikzpicture}[thick,every node/.style={transform shape},scale=0.70]

\filldraw (0,0) circle [radius=2pt] node (s1) [] {};
\filldraw (4,0) circle [radius=2pt] node (s2) [] {};

\path (s1) edge[-, dashed] node[auto] {} (s2);

\node () at (0,0.3) {$s_1$};
\node () at (4,0.3) {$s_2$};

\node (A1) at (0,-0.5) {$\strat^1_\coA$};
\node (A2) at (4,-0.5) {$\strat^2_\coA$};

\path (A1) edge[->] node[auto] {} (A2);

\draw (0,-2) ellipse [x radius=15pt, y radius=12pt];
\draw (4,-2) ellipse [x radius=15pt, y radius=12pt];

\node () at (-1.5,-2) {$\Out[s_1,\strat^1_\coA]$};
\node () at (5.5,-2) {$\Out[s_2,\strat^2_\coA]$};

\filldraw (0,-2) circle [radius=1pt] node (u1) [] {};
\filldraw (4,-2) circle [radius=1pt] node (u2) [] {};

\node () at (0,-1.8) {\small $u_1$};
\node () at (4,-1.8) {\small $u_2$};

\path (u1) edge[<-, dashed] node[auto] {} (u2);

\node (B1) at (0,-3.6) {$\strat^1_\coB$};
\node (B2) at (4,-3.6) {$\strat^2_\coB$};

\path (B1) edge[->] node[auto] {} (B2);

\draw (0,-5) ellipse [x radius=15pt, y radius=12pt];
\draw (4,-5) ellipse [x radius=15pt, y radius=12pt];

\node () at (-1.9,-5) {$\Out[s_1,\strat^1_\coA \uplus \strat^1_\coB]$};
\node () at (5.9,-5) {$\Out [s_2,\strat^2_\coA \uplus \strat^2_\coB]$};

\filldraw (0,-5) circle [radius=1pt] node (v1) [] {};
\filldraw (4,-5) circle [radius=1pt] node (v2) [] {};

\node () at (0,-4.8) {\small $v_1$};
\node () at (4,-4.8) {\small $v_2$};

\path (v1) edge[<-, dashed] node[auto] {} (v2);

\draw[] (-4,1) rectangle (8,-6.7);
\draw[dotted] (-3.7,-1) rectangle (7.7,-6.4);
\draw[dotted] (-3.4,-1.3) rectangle (7.4,-2.7);
\draw[dotted] (-3.4,-3.0) rectangle (7.4,-6.1);
\draw[dotted] (-3.1,-4.2) rectangle (7.1,-5.8);

\end{tikzpicture}
\caption{The \textbf{$\coA$-Forth$_\mathsf{c}$} half of \ocsrAB-bisimulation}
\label{fig-bis-c}
\end{center}
\end{figure}

\begin{description}
\itemsep = 1pt

\item[\textbf{$\coA$-Forth$_\mathsf{c}$:}] For every joint action $\strat^{1}_{\coA}$ of $\coA$ at $s_1$ there is a joint action $\strat^{2}_{\coA}$ of $\coA$ at $s_2$, such that:

\begin{description}
\itemsep = 0pt

\item[\textbf{$\coA$-LocalBack$_\mathsf{c}$:}] For every $u_{2} \in \Out[s_{2},\strat^{2}_{\coA}]$ there exists $u_{1} \in \Out[s_{1},\strat^{1}_{\coA}]$ such that $u_1 \bisim u_2$.

\item[\textbf{$\coB$-Forth$_\mathsf{c}$:}] For every joint action $\strat^{1}_{\coB}$ of $\coB$ at $s_1$ there is a joint action $\strat^{2}_{\coB}$ of $\coB$ at $s_2$, such that:

\begin{description}
\itemsep = 0pt
\item[\textbf{($\coA \uplus \coB$)-LocalBack$_\mathsf{c}$:}] For every $u_{2} \in \Out[s_{2},\strat^{2}_{\coA} \uplus \strat^{2}_{\coB}]$ there exists $u_{1} \in \Out[s_{1},\strat^{1}_{\coA} \uplus \strat^{1}_{\coB}]$ such that $u_1 \bisim u_2$.
\end{description}
\end{description}

\item[\textbf{$\coA$-Back$_\mathsf{c}$:}] Like \textbf{$\coA$-Forth$_c$}, but with $1$ and $2$ swapped.
\end{description}

\item[\textbf{\ocsrdd-bisimulation:}] ~(For illustration, see Figure \ref{fig-bis-dd}.)

\begin{figure}[ht]
\begin{center}
\begin{tikzpicture}[thick,every node/.style={transform shape},scale=0.75]

\filldraw (0,0) circle [radius=2pt] node (s1) [] {};
\filldraw (4,0) circle [radius=2pt] node (s2) [] {};

\path (s1) edge[-, dashed] node[auto] {} (s2);

\node () at (0,0.3) {$s_1$};
\node () at (4,0.3) {$s_2$};

\node (B1) at (0,-0.5) {$\strat^1_\coB$};
\node (B2) at (4,-0.5) {$\strat^2_\coB$};

\path (B1) edge[->] node[auto] {} (B2);

\node (A1) at (0,-1.8) {$\strat^1_\coA$};
\node (A2) at (4,-1.8) {$\strat^2_\coA$};

\path (A1) edge[<-] node[auto] {} (A2);

\draw (0,-3.5) ellipse [x radius=15pt, y radius=12pt];
\draw (4,-3.5) ellipse [x radius=15pt, y radius=12pt];

\node () at (-1.5,-3.5) {$\Out[s_1,\strat^1_\coA]$};
\node () at (5.5,-3.5) {$\Out[s_2,\strat^2_\coA]$};

\filldraw (0,-3.5) circle [radius=1pt] node (u1) [] {};
\filldraw (4,-3.5) circle [radius=1pt] node (u2) [] {};

\node () at (0,-3.3) {\small $u_1$};
\node () at (4,-3.3) {\small $u_2$};

\path (u1) edge[->, dashed] node[auto] {} (u2);

\draw (0,-5.1) ellipse [x radius=15pt, y radius=12pt];
\draw (4,-5.1) ellipse [x radius=15pt, y radius=12pt];

\node () at (-1.9,-5.1) {$\Out[s_1,\strat^1_\coA \uplus \strat^1_\coB]$};
\node () at (5.9,-5.1) {$\Out[s_2,\strat^2_\coA \uplus \strat^2_\coB]$};

\filldraw (0,-5.1) circle [radius=1pt] node (v1) [] {};
\filldraw (4,-5.1) circle [radius=1pt] node (v2) [] {};

\node () at (0,-4.9) {\small $v_1$};
\node () at (4,-4.9) {\small $v_2$};

\path (v1) edge[<-, dashed] node[auto] {} (v2);

\draw[] (-4,1) rectangle (8,-6.7);
\draw[dotted] (-3.7,-1.1) rectangle (7.7,-6.4);
\draw[dotted] (-3.4,-2.5) rectangle (7.4,-6.1);
\draw[dotted] (-3.1,-2.8) rectangle (7.1,-4.1);
\draw[dotted] (-3.1,-4.4) rectangle (7.1,-5.8);

\end{tikzpicture}
\caption{The \textbf{$\coB$-Forth$_{\alpha}$} half of \ocsrdd-bisimulation}
\label{fig-bis-dd} 
\end{center}
\end{figure}
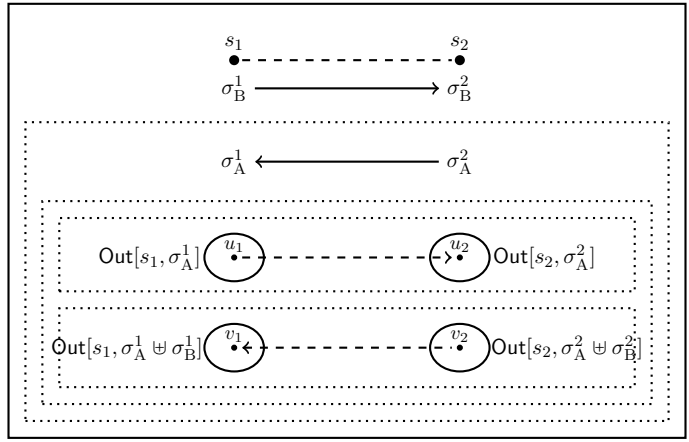

\begin{description}
\itemsep = 1pt

\item[\textbf{$\coB$-Forth$_{\alpha}$:}] For every joint action $\strat^{1}_{\coB}$ of $\coB$ at $s_1$ there is a joint action $\strat^{2}_{\coB}$ of $\coB$ at $s_2$, such that:

\begin{description}
\itemsep = 0pt

\item[\textbf{$\coA$-Back$_{\alpha}$:}] For every joint action $\strat^{2}_{\coA}$ of $\coA$ at $s_2$ there is a joint action $\strat^{1}_{\coA}$ of $\coA$ at $s_1$, such that: 

\begin{description}
\itemsep = 0pt

\item[\textbf{($\coA$)-LocalForth$_{\alpha}$:}] For every $u_{1} \in \Out[s_{1},\strat^{1}_{\coA}]$ there exists $u_{2} \in \Out[s_{2},\strat^{2}_{\coA}]$ such that $u_1 \bisim u_2$.

\item[\textbf{($\coA \uplus \coB$)-LocalBack$_{\alpha}$:}]
For every $u_{2} \in \Out[s_{2},\strat^{2}_{\coA} \uplus \strat^{2}_{\coB}]$ there exists $u_{1} \in \Out[s_{1},\strat^{1}_{\coA} \uplus \strat^{1}_{\coB}]$ such that $u_1 \bisim u_2$.
\end{description}
\end{description}

\item[\textbf{$\coB$-Back$_{\alpha}$:}] Like \textbf{$\coB$-Forth}, but with $1$ and $2$ swapped.
\end{description}

\item[\textbf{\ocsrdr-bisimulation:}] ~ (For illustration, see Figure \ref{fig-bis-dr}.)
 
\begin{figure}[h]
\begin{center}
\begin{tikzpicture}[thick,every node/.style={transform shape},scale=0.70]

\filldraw (0,0) circle [radius=2pt] node (s1) [] {};
\filldraw (4,0) circle [radius=2pt] node (s2) [] {};

\path (s1) edge[-, dashed] node[auto] {} (s2);

\node () at (0,0.3) {$s_1$};
\node () at (4,0.3) {$s_2$};

\node (A1) at (0,-0.5) {$\strat^1_\coA$};
\node (A2) at (4,-0.5) {$\strat^2_\coA$};

\path (A1) edge[->] node[auto] {} (A2);

\draw (0,-2) ellipse [x radius=15pt, y radius=12pt];
\draw (4,-2) ellipse [x radius=15pt, y radius=12pt];

\node () at (-1.5,-2) {$\Out[s_1,\strat^1_\coA]$};
\node () at (5.5,-2) {$\Out[s_2,\strat^2_\coA]$};

\filldraw (0,-2) circle [radius=1pt] node (u1) [] {};
\filldraw (4,-2) circle [radius=1pt] node (u2) [] {};

\node () at (0,-1.8) {\small $u_1$};
\node () at (4,-1.8) {\small $u_2$};

\path (u1) edge[<-, dashed] node[auto] {} (u2);

\node (B1) at (0,-3.6) {$\strat^1_\coB$};
\node (B2) at (4,-3.6) {$\strat^2_\coB$};

\path (B1) edge[<-] node[auto] {} (B2);

\draw (0,-5) ellipse [x radius=15pt, y radius=12pt];
\draw (4,-5) ellipse [x radius=15pt, y radius=12pt];

\node () at (-1.9,-5) {$\Out[s_1,\strat^1_\coA \uplus \strat^1_\coB]$};
\node () at (5.9,-5) {$\Out[s_2,\strat^2_\coA \uplus \strat^2_\coB]$};

\filldraw (0,-5) circle [radius=1pt] node (v1) [] {};
\filldraw (4,-5) circle [radius=1pt] node (v2) [] {};

\node () at (0,-4.8) {\small $v_1$};
\node () at (4,-4.8) {\small $v_2$};

\path (v1) edge[->, dashed] node[auto] {} (v2);

\draw[] (-4,1) rectangle (8,-6.7);
\draw[dotted] (-3.7,-1) rectangle (7.7,-6.4);
\draw[dotted] (-3.4,-1.3) rectangle (7.4,-2.7);
\draw[dotted] (-3.4,-3.0) rectangle (7.4,-6.1);
\draw[dotted] (-3.1,-4.2) rectangle (7.1,-5.8);

\end{tikzpicture}
\caption{The \textbf{$\coA$-Forth$_{\beta}$} half of \ocsrdr-bisimulation}
\label{fig-bis-dr} 
\end{center}
\end{figure}

\begin{description}
\itemsep = 1pt

\item[\textbf{$\coA$-Forth$_{\beta}$:}] For every joint action $\strat^{1}_{\coA}$ of $\coA$ at $s_1$ there is a joint action $\strat^{2}_{\coA}$ of $\coA$ at $s_2$, such that:

\begin{description}
\itemsep = 0pt

\item[\textbf{$\coA$-LocalBack$_{\beta}$:}] For every $u_{2} \in \Out[s_{2},\strat^{2}_{\coA}]$ there exists $u_{1} \in \Out[s_{1},\strat^{1}_{\coA}]$ such that $u_1 \bisim u_2$.

\item[\textbf{$\coB$-Back$_{\beta}$:}] For every joint action $\strat^{2}_{\coB}$ of $\coB$ at $s_2$ there is a joint action $\strat^{1}_{\coB}$ of $\coB$ at $s_1$, such that:

\begin{description}
\itemsep = 0pt 
\item[\textbf{($\coA \uplus \coB$)-LocalForth$_{\beta}$:}]
For every $u_{1} \in \Out[s_{1},\strat^{1}_{\coA} \uplus \strat^{1}_{\coB}]$ there exists $u_{2} \in \Out[s_{2},\strat^{2}_{\coA} \uplus \strat^{2}_{\coB}]$ such that $u_1 \bisim u_2$.
\end{description}
\end{description}

\item[\textbf{$\coA$-Back$_{\beta}$:}] Like \textbf{$\coA$-Forth}, but with $1$ and $2$ swapped.
\end{description}

\end{description}

\noindent States $s_1, s_2 \in \gmod$  are \defstyle{\CSR-bisimulation equivalent}, or just \defstyle{\CSR-bisimilar}
if there is a \CSR-bisimulation $\bisim$ in $\gmod$ such that $s_1 \bisim s_2$.
\end{definition}

\begin{proposition}[\CSR-bisimulation invariance]
\label{prop:CSRbisInv}
Let $\bisim$ be a \CSR-bisimulation in a game model $\gmod$. Then for every \CSR-formula $\theta$ and a pair $s_1, s_2 \in \gmod$ such that $s_1 \bisim s_2$: $\gmod, s_{1} \models \theta \ \mbox{iff} \ \gmod, s_{2} \models \theta$.
\end{proposition}

\begin{proof}
Induction on $\theta$. All boolean cases are straightforward. The cases for the 3 strategic operators are similar, but we will nevertheless check each of them, to ensure that the bisimulation conditions above are correctly defined.

\medskip

\noindent \textbf{(Case \ocsrAB)} Let $\theta = \condABd{\coA}{\coB}{\phi}{\psi}$, assuming that the claim holds for $\phi$ and $\psi$.
 
Suppose,  $\gmod, s_{1} \models \theta$. Then $\coA$ has a joint action $\strat^{1}_\coA$ at  $s_1$ such that, when applied, it guarantees $\phi$ and enables $\coB$ to adopt a joint action $\strat_\coB$ that is consistent with $\strat_\coA$ and guarantees $\psi$ when additionally applied by $\coB$.
By \textbf{$\coA$-Forth}$_\mathsf{c}$, there is a joint action $\strat^{2}_{\coA}$ of $\coA$ at $s_2$, such that, by \textbf{$\coA$-LocalBack}$_\mathsf{c}$, for each $u_{2} \in \Out[s_{2},\strat^{2}_{\coA}]$ there exists $u_{1} \in \Out[s_{1},\strat^{1}_{\coA}]$ such that $u_1 \bisim u_2$.
By the choice of $\strat^{1}_\coA$, $\gmod, u_{1} \models \phi$ for each $u_{1} \in \Out[s_{1},\strat^{1}_{\coA}]$.
It follows, by the inductive hypothesis applied to $\phi$, that $\gmod, u_{2} \models \phi$ for each $u_{2} \in \Out[s_{2},\strat^{2}_{\coA}]$.
Moreover, $\coB$ has a joint action $\strat^{1}_\coB$ at $s_1$ such that, when applied by $\coB$, in addition to $\coA$ applying $\strat^{1}_\coA$, it guarantees $\psi$, i.e. $\gmod, u_{1} \models \psi$ for each $u_{1} \in \Out[s_{1},\strat^{1}_{\coA} \uplus \strat^{1}_{\coB}]$.
By condition \textbf{$\coB$-Forth}$_\mathsf{c}$,  there is a joint action $\strat^{2}_{\coB}$ of $\coB$ at $s_2$, such that,
by (\textbf{$\coA \uplus \coB$)-LocalBack}$_\mathsf{c}$, for every $u_{2} \in \Out[s_{2},\strat^{2}_{\coA} \uplus \strat^{2}_{\coB}]$ there exists $u_{1} \in \Out[s_{1},\strat^{1}_{\coA} \uplus \strat^{1}_{\coB}]$ such that $u_1 \bisim u_2$.
Therefore, by the inductive hypothesis applied to $\psi$, $\gmod, u_{2} \models \psi$ for each $u_{2} \in \Out[s_{2},\strat^{2}_{\coA} \uplus \strat^{2}_{\coB}]$. Thus, $\gmod, s_{2} \models \theta$.

The converse is similar, using \textbf{$\coA$-Back}$_\mathsf{c}$.

\medskip

\noindent \textbf{(Case \ocsrdd)} Let $\theta = \dedic{\coA}{\coB}{\phi}{\psi}$, assuming the claim holds for $\phi$ and $\psi$.

Suppose,  $\gmod, s_{1} \models \theta$. Let $\strat^{1}_{\coB}$ be a joint action of $\coB$ at $s_1$ satisfying the truth condition of $\theta$. By \textbf{$\coB$-Forth$_{\alpha}$}, there is a joint action $\strat^{2}_{\coB}$ of $\coB$ at $s_2$, such that \textbf{$\coA$-Back$_{\alpha}$} holds.
Now, take any joint action $\strat^{2}_{\coA}$ of $\coA$ at $s_2$ such that $\gmod, u_{2} \models \phi$ for each $u_{2} \in \Out[s_{2},\strat^{2}_{\coA}]$.
Then, by \textbf{$\coA$-Back$_{\alpha}$}, there is a joint action $\strat^{1}_{\coA}$ of $\coA$ at $s_1$ such that, by \textbf{($\coA$)-LocalForth$_{\alpha}$}, for every $u_{1} \in \Out[s_{1},\strat^{1}_{\coA}]$ there exists $u_{2} \in \Out[s_{2},\strat^{2}_{\coA}]$ such that $u_1 \bisim u_2$. Then, by the inductive hypothesis applied to $\phi$, it follows that $\gmod, u_{1} \models \phi$ for each $u_{1} \in \Out[s_{1},\strat^{1}_{\coA}]$. By the choice of $\strat^{1}_{\coB}$, this implies $\gmod, u_{1} \models \psi$ for each $u_{1} \in \Out[s_{1},\strat^{1}_{\coA} \uplus \strat^{1}_{\coB}]$.
By (\textbf{$\coA \uplus \coB$)-LocalBack$_{\alpha}$}, for every $u_{2} \in \Out[s_{2},\strat^{2}_{\coA} \uplus \strat^{2}_{\coB}]$ there exists $u_{1} \in \Out[s_{1},\strat^{1}_{\coA} \uplus \strat^{1}_{\coB}]$ such that $u_1 \bisim u_2$.
Therefore, by the inductive hypothesis applied to $\psi$, we have $\gmod, u_{2} \models \psi$ for each $u_{2} \in \Out[s_{2},\strat^{2}_{\coA} \uplus \strat^{2}_{\coB}]$. Thus, $\strat^{2}_{\coB}$ satisfies the truth condition of $\theta$ at $s_{2}$. Hence, $\gmod, s_{2} \models \theta$.

The converse direction is analogous, using \textbf{$\coB$-Back}$_{\alpha}$.

\medskip

\noindent \textbf{(Case \ocsrdr)} Let $\theta = \dere{\coA}{\coB}{\phi}{\psi}$, assuming the claim holds for $\phi$ and $\psi$.

Suppose,  $\gmod, s_{2} \models \theta$.
Then, consider any joint action $\strat^{1}_\coA$ of  $\coA$ at $s_1$ such that, when applied by $\coA$, it guarantees $\phi$. (If no such joint action exist at  $s_1$, then $\gmod, s_{1} \models \theta$ is vacuously true.)
Then, by \textbf{$\coA$-Forth$_{\beta}$}, there is a joint action $\strat^{2}_{\coA}$ of $\coA$ at $s_2$, such that, by \textbf{$\coA$-LocalBack$_{\beta}$}, for every $u_{2} \in \Out[s_{2},\strat^{2}_{\coA}]$ there exists $u_{1} \in \Out[s_{1},\strat^{1}_{\coA}]$ such that $u_1 \bisim u_2$.
Then, by the inductive hypothesis applied to $\phi$, it follows that $\gmod, u_{2} \models \phi$ for each $u_{2} \in \Out[s_{2},\strat^{2}_{\coA}]$. Therefore, by the assumption for the truth of $\theta$ at $s_{2}$, $\coB$ has a joint action $\strat^{2}_\coB$ at $s_2$ such that, when applied by $\coB$, in addition to $\coA$ applying $\strat^{2}_\coA$, it guarantees the truth of $\psi$, i.e. $\gmod, u_{2} \models \psi$ for each $u_{2} \in \Out[s_{2},\strat^{2}_{\coA} \uplus \strat^{2}_{\coB}]$.
Then, by \textbf{$\coB$-Back$_{\beta}$}, there is a joint action $\strat^{1}_{\coB}$ of $\coB$ at $s_1$, such that, by (\textbf{$\coA \uplus \coB$)-LocalForth$_{\beta}$}, for every $u_{1} \in \Out[s_{1},\strat^{1}_{\coA} \uplus \strat^{1}_{\coB}]$ there exists $u_{2} \in \Out[s_{2},\strat^{2}_{\coA} \uplus \strat^{2}_{\coB}]$ such that $u_1 \bisim u_2$.
Therefore, $\gmod, u_{1} \models \psi$ for each $u_{1} \in \Out[s_{1},\strat^{1}_{\coA} \uplus \strat^{1}_{\coB}]$. Thus, $\gmod, s_{1} \models \theta$.

The converse is analogous, using \textbf{$\coA$-Back}$_{\beta}$.
\end{proof}

We also obtain the Hennessy-Milner property for \CSR-bisimulations:

\begin{proposition}[Hennessy-Milner property]
\label{prop:CSRbisHM}
For any finite game model $\gmod$ the relation of $\CSR$-equivalence
(satisfaction of the same $\CSR$-formulae) between states in $\gmod$ is
a \CSR-bisimulation in $\gmod$.
\end{proposition}

\begin{proof}
(Sketch) One direction follows from Prop. \ref{prop:CSRbisInv}. We now prove the converse.
Since $\gmod$ is finite, there is a mapping $\chi$ from $\gmod$ to the formulae of $\CSR$ that assigns to each state $s$ in $\gmod$ its characteristic formula $\chi(s)$, such that $s_1,s_2$ are $\CSR$-equivalent if and only if $s_1$ satisfies $\chi(s_2)$ (and vice versa), iff $\chi(s_1) \equiv \chi(s_2)$. Furthermore, $\chi(s_1) \wedge \chi(s_2) \equiv \bot$ whenever $s_1$ and $s_2$ are not $\CSR$-equivalent.
Now, for any set of states $Z$ in $\gmod$ we define $\chi(Z)  := \bigvee_{z\in Z} \chi(z)$.
The crucial observation for proving the claim is that every state $s\in \gmod$ satisfies each of the following formulae, enabling the verification of the respective $\CSR$-bisimulation conditions:

{\small
\[
\textbf{(1)}
\bigwedge_{A,B \subseteq \Agt} \big\{
\condABd{\coA}{\coB}{\chi(Z)}{\chi(Y)}
\mid
\exists \sigma \in \Act_s:
 \Out[s,\sigma\vert_A] = Z
\mbox{ and }
 \Out[s,\sigma\vert_{(A\cup B)}] = Y
\big\}
\]
\vspace{1mm}
\[
\textbf{(2)}
\bigwedge_{A,B \subseteq \Agt} \big\{
\dedic{\coA}{\coB}{\chi(Z)}{\chi(Y)}
\mid
\exists \sigma \in \Act_s:
~ 
\hspace{46mm}
\]
\[
\forall \sigma' \in \Act_s
\mbox{ if }
\Out[s,\sigma'\vert_A] \subseteq Z
\mbox{ and }
\sigma' \vert_{(\mathrm{B\setminus A})} =  \sigma \vert_{(\mathrm{B\setminus A})}
\mbox{ then }
 \Out[s,\sigma' \vert_{(A\cup B)}] \subseteq Y
\big\}
\]
\vspace{1mm}
\[
\textbf{(3)}
\bigwedge_{A,B \subseteq \Agt} \big\{
\dere{\coA}{\coB}{\chi(Z)}{\chi(Y)}
\mid \forall \sigma \in \Act_s:
~
\hspace{46mm}
\]
\[
  \Out[s,\sigma\vert_A] \subseteq Z
\mbox{ implies }
 \Out[s,\sigma' \vert_{(A\cup B)}] \subseteq Y
\mbox{ for some } \sigma' \in \Act_s
\mbox{ such that }  \sigma' \vert_{\coA} =  \sigma \vert_{\coA}
\big\}
\]
}
\end{proof}

\subsection{Some remarks on expressiveness and definability}

\begin{proposition}
\label{prop:expressiveness}
Let $\aga, \agb$ be different agents and $p,q$ be different atomic propositions. Then the following hold\footnote{Even though we state the non-definability claims for \CL, they apply likewise even to \ATLs with its standard, memory-based semantics, because all formulae of \ATLs are invariant under \CL-bisimulations with respect to that semantics.}.

\begin{enumerate}
\itemsep = 2pt

\item $\dere{\aga}{\agb}{p}{q} \not \equiv \dedic{\aga}{\agb}{p}{q}$.

\item $\condABd{\aga}{\agb}{p}{q}$ is not definable in $\CL$.
\item $\condb{\aga}{p}{q}$ (and, consequently, $\dere{\aga}{\emptyset}{p}{q}$) is not definable in $\CL$.
\item $\dedic{\agb}{\aga}{q}{p}$ is not definable in $\CL$.
\end{enumerate}
\end{proposition}

\begin{proof}
The claims follow respectively from examples \ref{example2}, \ref{exampleA}, \ref{exampleB} and \ref{exampleC}.
\end{proof}

\noindent The results above generalise to pairwise coalitions in a straightforward way.

\newpage
\begin{example}
\label{exampleA}

The game models $\M_1$, $\M_2$ below involve three players: $\aga$, $\agb$, $\agc$.

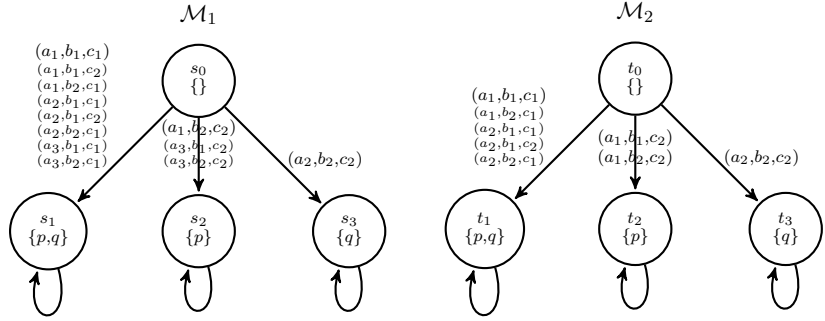
\begin{center}
\begin{tabular}{cc}

\begin{tikzpicture}[->,>=stealth',shorten >=1pt,node distance=25mm,thick,every node/.style={transform shape},scale=0.8]
\tikzstyle{every state}=[draw=black,text=black,minimum size=12mm]

\node[state] (0) {$s_0 \atop \{\}$};
\node[state] (2) [below of=0] {$s_2 \atop \{p\}$};
\node[state] (1) [left of=2] {$s_1 \atop \{p, q\}$};
\node[state] (3) [right of=2] {$s_3 \atop \{q\}$};

\path
(1) edge [loop below] node {} (1)
(2) edge [loop below] node {} (2)
(3) edge [loop below] node {} (3)
 
(0) edge [] node {} (1)
(0) edge [right] node {} (2)
(0) edge [right] node {} (3);

\node [above=8mm] at (0) {$\M_1$};
\node [above,xshift=4mm,yshift=9mm] at (1) {$(a_1, b_1, c_1) \atop {(a_1, b_1, c_2) \atop {(a_1, b_2, c_1) \atop {(a_2, b_1, c_1) \atop {(a_2, b_1, c_2) \atop {(a_2, b_2, c_1) \atop {(a_3, b_1, c_1) \atop (a_3, b_2, c_1)}}}}}}$};
\node [above,xshift=0mm,yshift=9mm] at (2) {$(a_1, b_2, c_2) \atop {(a_3, b_1, c_2) \atop (a_3, b_2, c_2)}$};
\node [above,xshift=-4mm,yshift=9mm] at (3) {$\atop (a_2, b_2, c_2)$};

\end{tikzpicture}

\hspace{5mm}

\begin{tikzpicture}[->,>=stealth',shorten >=1pt,node distance=25mm,thick,every node/.style={transform shape},scale=0.8]
\tikzstyle{every state}=[draw=black,text=black,minimum size=12mm]

\node[state] (0) {$t_0 \atop \{\}$};
\node[state] (2) [below of=0] {$t_2 \atop \{p\}$};
\node[state] (1) [left of=2] {$t_1 \atop \{p, q\}$};
\node[state] (3) [right of=2] {$t_3 \atop \{q\}$};

\path
(1) edge [loop below] node {} (1)
(2) edge [loop below] node {} (2)
(3) edge [loop below] node {} (3)
 
(0) edge [] node {} (1)
(0) edge [] node {} (2)
(0) edge [] node {} (3);

\node [above=8mm] at (0) {$\M_2$};
\node [above,xshift=4mm,yshift=9mm] at (1) {$(a_1, b_1, c_1) \atop {(a_1, b_2, c_1) \atop {(a_2, b_1, c_1) \atop {(a_2, b_1, c_2) \atop (a_2, b_2, c_1)}}}$};
\node [above,xshift=0mm,yshift=9mm] at (2) {$(a_1, b_1, c_2) \atop (a_1, b_2, c_2)$};
\node [above,xshift=-4mm,yshift=9mm] at (3) {$\atop (a_2, b_2, c_2)$};

\end{tikzpicture}
\end{tabular}
\end{center}

\noindent Note that (1) The relation $\bisim = \{(s_i,t_i) \mid i = 0,1,2,3\}$ is an \CL-bisimulation between $\M_1$ and $\M_2$; (2) $\M_1, s_0 \Vdash \condABd{\aga}{\agb}{p}{q}$ but $\M_2, t_0 \not \Vdash \condABd{\aga}{\agb}{p}{q}$.
\end{example}

\begin{example}
\label{exampleB}

The game models $\M_1$ and $\M_2$  below involve two players: $\aga$ and $\agb$.

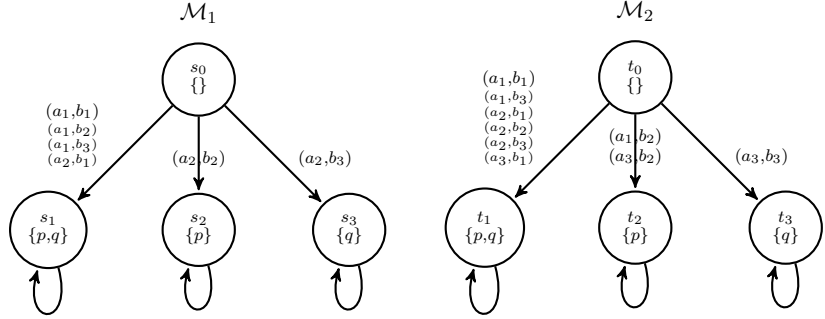
\begin{center}
\begin{tabular}{cc}

\begin{tikzpicture}[->,>=stealth',shorten >=1pt,node distance=25mm,thick,every node/.style={transform shape},scale=0.8]
\tikzstyle{every state}=[draw=black,text=black,minimum size=12mm]

\node[state] (0) {$s_0 \atop \{\}$};
\node[state] (2) [below of=0] {$s_2 \atop \{p\}$};
\node[state] (1) [left of=2] {$s_1 \atop \{p, q\}$};
\node[state] (3) [right of=2] {$s_3 \atop \{q\}$};

\path
(1) edge [loop below] node {} (1)
(2) edge [loop below] node {} (2)
(3) edge [loop below] node {} (3)
 
(0) edge [above,xshift=-2mm,yshift=7mm] node {} (1)
(0) edge [above,xshift=0mm,yshift=9mm] node {} (2)
(0) edge [above,xshift=-4mm,yshift=9mm] node {} (3);

\node [above=8mm] at (0) {$\M_1$};
\node [above,xshift=4mm,yshift=9mm] at (1) {$(a_1, b_1) \atop {(a_1, b_2) \atop {(a_1, b_3) \atop (a_2, b_1)}}$};
\node [above,xshift=0mm,yshift=9mm] at (2) {$\atop (a_2, b_2)$};
\node [above,xshift=-4mm,yshift=9mm] at (3) {$\atop (a_2, b_3)$};

\end{tikzpicture}

\hspace{5mm}

\begin{tikzpicture}[->,>=stealth',shorten >=1pt,node distance=25mm,thick,every node/.style={transform shape},scale=0.8]
\tikzstyle{every state}=[draw=black,text=black,minimum size=12mm]

\node[state] (0) {$t_0 \atop \{\}$};
\node[state] (2) [below of=0] {$t_2 \atop \{p\}$};
\node[state] (1) [left of=2] {$t_1 \atop \{p, q\}$};
\node[state] (3) [right of=2] {$t_3 \atop \{q\}$};
\path
(1) edge [loop below] node {} (1)
(2) edge [loop below] node {} (2)
(3) edge [loop below] node {} (3)
 
(0) edge [] node {} (1)
(0) edge [] node {} (2)
(0) edge [] node {} (3);

\node [above=8mm] at (0) {$\M_2$};
\node [above,xshift=4mm,yshift=9mm] at (1) {$(a_1, b_1) \atop {(a_1, b_3) \atop {(a_2, b_1) \atop {(a_2, b_2) \atop {(a_2, b_3) \atop (a_3, b_1)}}}}$};
\node [above,xshift=0mm,yshift=9mm] at (2) {$(a_1, b_2) \atop (a_3, b_2)$};
\node [above,xshift=-4mm,yshift=9mm] at (3) {$\atop (a_3, b_3)$};

\end{tikzpicture}
\end{tabular}
\end{center}

\noindent Note that  (1) The relation $\bisim = \{(s_i,t_i) \mid i = 0,1,2,3\}$ is an \CL-bisimulation between $\M_1$ and $\M_2$; (2) $\M_1, s_0 \Vdash \condb{\aga}{p}{q}$ but $\M_2, t_0 \not \Vdash \condb{\aga}{p}{q}$.
\end{example}

\begin{example}
\label{exampleC}

The game models $\M_1$, $\M_2$  below involve three players: $\aga$, $\agb$, $\agc$.

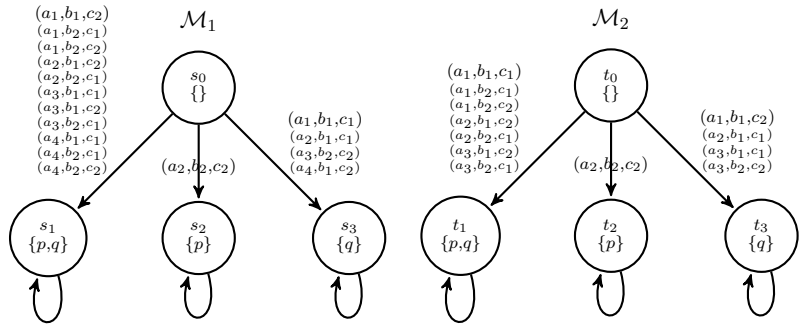
\begin{center}
\begin{tabular}{cc}

\begin{tikzpicture}[->,>=stealth',shorten >=1pt,node distance=25mm,thick,every node/.style={transform shape},scale=0.8]
\tikzstyle{every state}=[draw=black,text=black,minimum size=12mm]

\node[state] (0) {$s_0 \atop \{\}$};
\node[state] (2) [below of=0] {$s_2 \atop \{p\}$};
\node[state] (1) [left of=2] {$s_1 \atop \{p, q\}$};
\node[state] (3) [right of=2] {$s_3 \atop \{q\}$};

\path
(1) edge [loop below] node {} (1)
(2) edge [loop below] node {} (2)
(3) edge [loop below] node {} (3)
 
(0) edge [] node {} (1)
(0) edge [] node {} (2)
(0) edge [] node {} (3);

\node [above=8mm] at (0) {$\M_1$};
\node [above,xshift=4mm,yshift=9mm] at (1) {$(a_1, b_1, c_2) \atop {(a_1, b_2, c_1) \atop {(a_1, b_2, c_2) \atop {(a_2, b_1, c_2) \atop {(a_2, b_2, c_1) \atop {(a_3, b_1, c_1) \atop {(a_3, b_1, c_2) \atop {(a_3, b_2, c_1) \atop {(a_4, b_1, c_1) \atop {(a_4, b_2, c_1) \atop (a_4, b_2, c_2)}}}}}}}}}$};
\node [above,xshift=0mm,yshift=9mm] at (2) {$\atop (a_2, b_2, c_2)$};
\node [above,xshift=-4mm,yshift=9mm] at (3) {$(a_1, b_1, c_1) \atop {(a_2, b_1, c_1) \atop {(a_3, b_2, c_2) \atop (a_4, b_1, c_2)}}$};

\end{tikzpicture}

\hspace{5pt}

\begin{tikzpicture}[->,>=stealth',shorten >=1pt,node distance=25mm,thick,every node/.style={transform shape},scale=0.8]
\tikzstyle{every state}=[draw=black,text=black,minimum size=12mm]

\node[state] (0) {$t_0 \atop \{\}$};
\node[state] (2) [below of=0] {$t_2 \atop \{p\}$};
\node[state] (1) [left of=2] {$t_1 \atop \{p, q\}$};
\node[state] (3) [right of=2] {$t_3 \atop \{q\}$};

\path
(1) edge [loop below] node {} (1)
(2) edge [loop below] node {} (2)
(3) edge [loop below] node {} (3)
 
(0) edge [] node {} (1)
(0) edge [] node {} (2)
(0) edge [] node {} (3);

\node [above=8mm] at (0) {$\M_2$};
\node [above,xshift=4mm,yshift=9mm] at (1) {$(a_1, b_1, c_1) \atop {(a_1, b_2, c_1) \atop {(a_1, b_2, c_2) \atop {(a_2, b_1, c_2) \atop {(a_2, b_2, c_1) \atop {(a_3, b_1, c_2) \atop (a_3, b_2, c_1)}}}}}$};
\node [above,xshift=0mm,yshift=9mm] at (2) {$\atop (a_2, b_2, c_2)$};
\node [above,xshift=-4mm,yshift=9mm] at (3) {$(a_1, b_1, c_2) \atop {(a_2, b_1, c_1) \atop {(a_3, b_1, c_1) \atop (a_3, b_2, c_2)}}$};

\end{tikzpicture}
\end{tabular}
\end{center}

\noindent Note that (1) The relation $\bisim = \{(s_i,t_i) \mid i = 0,1,2,3\}$ is an \CL-bisimulation between $\M_1$ and $\M_2$; (2) $\M_1, s_0 \Vdash \dedic{\agb}{\aga}{q}{p}$ but $\M_2, t_0 \not \Vdash \dedic{\agb}{\aga}{q}{p}$.
\end{example}

\section{Axiomatic system for \CSR}  
\label{sec:Axioms}

Here we propose systems of axiom schemes for each of the basic operators of  \CSR and for the whole logic, without stating completeness claims; these are left for a follow-up work.
Some of these axiom schemes are adapted from the axiomatic systems for \textsf{SFCL} and \textsf{GPCL} presented in \cite{GorankoEnqvist18}. 

\subsection{Common axiom schemes and rules for \CSR}  
\label{subsec:CommonAxioms}

The axiomatic system $\Ax_{\CSR}$ builds on the axiom schemes and rules of the complete axiomatic system for Coalition Logic $\Ax_{\CL}$, given in \cite{Pauly01phd} and \cite{Pauly02modal}. 
Analogues of these axiom schemes and rules are added for each of the conditional strategic operators occurring in the fragment of \CSR that is to be axiomatized. 
(Recall that  the coalitional operator of \CL is a special case of each of \ocsrAB, \ocsrdd, \ocsrdr.) 
Some of these common axiom schemes and rules will turn out derivable from the special ones added below, but we are not concerned now with minimality of our system.

\subsection{Additional axiom schemes  and rules for $\ocsrAB$}  
\label{subsec:AxiomsOc}
%
\noindent \textbf{Axiom schemes:}
\begin{description}
\itemsep = 2pt

\item[(\ocsrAB 1)]
Monotonicity w.r.t. $\coA$:
  
$\condABd{\coA}{\coB}{\phi}{\psi} \to \condABd{\coA \cup \coC}{\coB}{\phi}{\psi}$ for any $\coC \subseteq \Agt$

\item[(\ocsrAB 2)]
Monotonicity w.r.t. $\coB$:

$\condABd{\coA}{\coB}{\phi}{\psi} \to \condABd{\coA}{\coB\cup \coC}{\phi}{\psi}$ for any $\coC \subseteq \Agt$

\item[(\ocsrAB 3)]
$\condABd{\coA}{\coB}{\phi}{\psi} \to \condABd{\coA\cup \coB}{\emptyset}{(\phi \land \psi)}{\top}$

\item[(\ocsrAB 4)]
$\condABd{\coA}{\emptyset}{\phi}{\psi} \ifff \condABd{\coA}{\emptyset}{(\phi\land \psi)}{\top}$

(NB: the direction $\to$ follows from (\ocsrAB 3).)

\item[(\ocsrAB 5)]    
$\lnot \condABd{\coA}{\coB}{\bot}{\psi}$

\item[(\ocsrAB 6)]
$\condABd{\coA}{\coB}{\phi}{\psi} \ifff \condABd{\coA}{\coB\!\setminus\!\coA}{\phi}{\,\psi}$

\item[(\ocsrAB 7)]
$\condABd{\coA}{\coB}{\phi}{\psi} \ifff \condABd{\coA}{\coB}{\phi}{(\phi \land \psi)}$

\end{description}
 
\smallskip           
\noindent \textbf{Rule of inference:} 
\textbf{$\ocsrAB$-Monotonicity (\ocsrAB-Mon)}: 
\[\frac{\phi \to \phi', \ \psi \to \psi'}{\condABd{\coA}{\coB}{\phi}{\psi} \to \condABd{\coA}{\coB}{\phi'}{\psi'}}\]

\subsection{Additional axiom schemes and rules for $\ocsrdr$} 
\label{subsec:AxiomsOdr}

\noindent \textbf{Axiom schemes:}
\begin{description}
\itemsep = 2pt

\item[(\ocsrdr 1)]
Monotonicity w.r.t. $\coB$:
  
$\dere{\coA}{\coB}{\phi}{\psi} \to \dere{\coA}{\coB\cup \coC}{\phi}{\psi}$ for any $\coC \subseteq \Agt$.
   
\item[(\ocsrdr 2)]
$\dere{\coA}{\emptyset}{\phi}{\phi}$

\item[(\ocsrdr 3)]
$\dere{\coA}{\emptyset}{\bot}{\psi}$

\item[(\ocsrdr 4)]
$\dere{\emptyset}{\coA}{\top}{\phi} \to \lnot \dere{\coA}{\coB}{\phi}{\bot}$

\item[(\ocsrdr 5)]
$\dere{\coA}{\coB}{\phi}{\psi} \ifff \dere{\coA}{\coB\!\setminus\!\coA}{\phi}{\,\psi}$
  
\item[(\ocsrdr 6)]  
$\dere{\coA}{\coB}{\phi}{\psi} \ifff \dere{\coA}{\coB}{\phi}{(\phi \land \psi)}$
\end{description}

\smallskip
\noindent \textbf{Rule of inference:}
\textbf{$\ocsrdr$-Monotonicity (\ocsrdr-Mon)}:
\[\frac{\phi' \to \phi, \  \psi \to \psi'}{\dere{\coA}{\coB}{\phi}{\psi} \to \dere{\coA}{\coB}{\phi'}{\psi'}}\]

\subsection{Additional axiom schemes and rules for $\ocsrdd$}
\label{subsec:AxiomsOdd}

\noindent \textbf{Axiom schemes:} 
All axioms (\ocsrdr 1) - (\ocsrdr 6), rewritten for $\ocsrdd$. In addition: 

\begin{description}
\itemsep = 0pt

\item[(\ocsrdd *)]
Anti-monotonicity w.r.t. $\coA$:

$\dedic{\coA\cup \coC}{\coB}{\phi}{\psi} \to \dedic{\coA}{\coB}{\phi}{\psi}$ for any $\coC \subseteq \Agt$.
\end{description}

\smallskip
\noindent \textbf{Rule of inference:}
\textbf{$\ocsrdd$-Monotonicity (\ocsrdd-Mon)}:
\[\frac{\phi' \to \phi, \  \psi \to \psi'}{\dedic{\coA}{\coB}{\phi}{\psi} \to \dedic{\coA}{\coB}{\phi'}{\psi'}}\]

\subsection{Interacting axioms for \CSR}  
\label{subsec:InterAxioms}

\begin{description}
\itemsep = 2pt

\item[(\CSR 1)]
$\dedic{\coA}{\coB}{\phi}{\psi} \to \dere{\coA}{\coB}{\phi}{\psi}$

\item[(\CSR 2)]
$\dere{\emptyset}{\coA}{\top}{\phi} \land \dere{\coA}{\coB}{\phi}{\psi} \to \condABd{\coA}{\coB}{\phi}{\psi}$
\footnote{An analogue of (\CSR 2) for $\ocsrdd$ is easily derivable from (\CSR 1) and (\CSR 2).} 
\end{description}

\begin{proposition}[Soundness] 
\label{prop:anti-monotonicity} 
The following hold for the system $\Ax_{\CSR}$.  
\begin{enumerate}
\item All axiom schemes are valid in the formal semantics of \CSR. 
\item 
The analogue of the anti-monotonicity axiom with respect to $\coA$ for $\ocsrdr$: 
  
$\dere{\coA\cup \coC}{\coB}{\phi}{\psi} \to \dere{\coA}{\coB}{\phi}{\psi}$, 
for any $\coA,\coB,\coC \subseteq \Agt$, 
is not valid. 
\end{enumerate}
\end{proposition}
\begin{proof}
Checking the soundness of most of the axioms is routine application of the formal semantics and we leave the details to the reader. 
We will only verify here the anti-monotonicity axiom scheme (\ocsrdd *),   
which is less straightforward and plays a special role of distinguishing \ocsrdr\, and \ocsrdd.

It suffices to prove the validity of the following instance:  

\smallskip
$\dedic{\coA\cup \coC}{\coB}{p}{q} \to \dedic{\coA}{\coB}{p}{q}$, 
for any $\coA,\coB,\coC \subseteq \Agt$.

\smallskip
Consider any CGM $\cgm$ and a state $s$ in it. 

For any formula $\phi$ we denote 
$\extn{\phi}_\cgm := \{ w \in \cgm \mid \cgm, w \Vdash \phi \}$. 

\smallskip
Suppose $\cgm, s \Vdash \dedic{\coA\cup \coC}{\coB}{p}{q}$.  
Fix a joint action $\strat_\coB$ witnessing the truth of that antecedent. 
Thus, for every joint action $\strat_{\coA\cup \coC}$ such that 
$\Out[s,\strat_{\coA\cup \coC}] \subseteq  \extn{p}_\cgm$ we have that 
$\Out[s,\strat_{(\coA\cup \coC)} \uplus \strat_{\coB}] \subseteq  \extn{q}_\cgm$.

To show that  $\cgm, s \Vdash \dedic{\coA}{\coB}{p}{q}$, we use the same  joint action $\strat_\coB$. 

Consider any  joint action $\strat_\coA$ such that 
 $\Out[s,\strat_{\coA}] \subseteq  \extn{p}_\cgm$. 
 
 That implies that for any additional 
 joint action $\strat_{\coC \setminus \coA}$, we have for the resulting joint action 
 $\strat_{\coA\cup \coC} :=  \strat_{\coA} \uplus \strat_{\coC \setminus \coA}$ that 
 $\Out[s,\strat_{\coA\cup \coC}] \subseteq  \extn{p}_\cgm$. 
 
 Then, by the assumption above, we have that 
 $\Out[s,\strat_{(\coA\cup \coC)} \uplus \strat_{\coB}] \subseteq  \extn{q}_\cgm$. 
 
Now, note that every extension of $\strat_{\coA} \uplus \strat_{\coB}$ to a full strategy profile $\strat'$ can be obtained by first selecting a joint action $\strat'_{\coC}$ which coincides with $\strat_{\coA} \uplus \strat_{\coB}$ on $\coA \cup \coB$ and is defined according to $\strat'$ for the agents in  $\coC  \setminus (\coA \cup \coB)$. 
Equivalently, every such strategy profile $\strat'$  can be generated 
by first selecting a joint action $\strat_{\coA\cup \coC}$ which extends $\strat_{\coA}$ with the joint action $\strat_{\coC \setminus \coA}$ obtained by restricting $\strat'_{\coC}$ to $\coC \setminus \coA$, then adding the actions of the agents from $\coB \setminus  (\coA\cup \coC)$ according to  $\strat_{\coA} \uplus \strat_{\coB}$, and then adding the remaining actions according to $\strat'$. Thus, $\strat'$ can be constructed as an extension of $\strat_{(\coA\cup \coC)} \uplus \strat_{\coB}$, where $\strat_{(\coA\cup \coC)}$ extends $\strat_\coA$.  
Hence, the outcome from $s$ of every such $\strat'$ is in $ \extn{q}_\cgm$.

Therefore $\Out[s,\strat_{\coA} \uplus \strat_{\coB}] \subseteq  \extn{q}_\cgm$. 
Hence,  $\cgm, s \Vdash \dedic{\coA}{\coB}{p}{q}$. 
Thus, we have shown that $\cgm, s \Vdash \dedic{\coA\cup \coC}{\coB}{p}{q}  \to \dedic{\coA}{\coB}{p}{q}$, whence the validity of that formula. 

\medskip
2. The instance $\dere{\{\aga, \agc\}}{\agb}{p}{q} \to \dere{\aga}{\agb}{p}{q}$ 
of the anti-monotonicity principle w.r.t. $\coA$ for $\ocsrdr$ is falsified in the example below of a game model $\M$ with three players, $\aga$, $\agb$ and $\agc$.

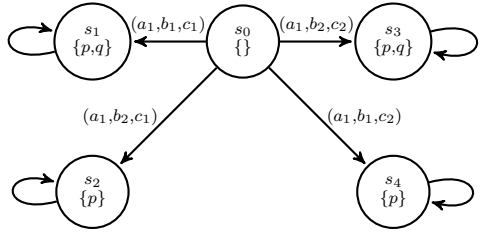
\begin{center}
\begin{tikzpicture}[->,>=stealth',shorten >=1pt,node distance=25mm,thick,every node/.style={transform shape},scale=0.8]
\tikzstyle{every state}=[draw=black,text=black,minimum size=12mm]

\node[state] (0) {$s_0 \atop \{\}$};
\node[state] (1) [left of=0] {$s_1 \atop \{p,q\}$};
\node[state] (2) [below of=1] {$s_2 \atop \{p\}$};
\node[state] (3) [right of=0] {$s_3 \atop \{p,q\}$};
\node[state] (4) [below of=3] {$s_4 \atop \{p\}$};

\path 
(1) edge [loop left] node {} (1)
(2) edge [loop left] node {} (2)
(3) edge [loop right] node {} (3)
(4) edge [loop right] node {} (4)

(0) edge [above] node {$_{({a_1}, b_1, c_1)}$} (1)
(0) edge [left] node {$_{({a_1}, {b_2}, c_1)}$} (2)
(0) edge [above] node {$_{({a_1}, b_2, c_2)}$} (3)  
(0) edge [right] node {$_{({a_1}, b_1, c_2)}$} (4);

\end{tikzpicture}
\end{center}

\noindent It is easy to verify that $\M, s_0 \Vdash \dere{\{\aga, \agc\}}{\agb}{p}{q}$ but $\M, s_0 \not \Vdash \dere{\aga}{\agb}{p}{q}$. 
Hence, 
$\M, s_0 \not \Vdash \dere{\{\aga, \agc\}}{\agb}{p}{q} \to \dere{\aga}{\agb}{p}{q}$.

\end{proof}

Thus, 
the anti-monotonicity principle with respect to $\coA$ is (at least so far) the only axiom scheme in our system, that distinguishes $\ocsrdd$ from $\ocsrdr$.

\section{Concluding remarks: the road ahead}
\label{sec:Concluding}

First, we note that, while the new strategic operators introduced here can be expressed in a suitable version of Strategy Logic (cf. \cite{corr/MogaveroMPV16}), we choose -- for both conceptual and computational reasons -- to stay within a purely modal framework where actions and strategies are not explicitly referred and quantified over in the language, but are only present in the semantics.

We regard this work as a step towards developing a rich technical framework for logic-based conditional strategic reasoning of rational agents. The major further steps and directions include:

\begin{enumerate}
\item Completeness proofs of the proposed axiomatizations of the 3 main fragments and the entire logic \CSR are currently under construction. 

\item We also claim that, for general reasons, \CSR has the finite tree-model property, and is therefore decidable (to be proved in a follow-up paper). 
We further conjecture that its satisfiability problem is PSPACE-complete -- a major argument in favour of the modal approach to formalising conditional strategic reasoning advocated here, as opposed to one based on a version of Strategy Logic. We also leave the question of the precise complexity of model checking subject to further investigation, but conjecture that it will still be tractable in the size of the model, as in \CL and \ATL.

\item Extension of the framework to a full-fledged, \emph{long term} conditional strategic reasoning, by extending the language with standard temporal operators, to produce an \ATL-like extension of \CSR.
 
\item Long term conditional strategic reasoning naturally requires considerations about strategic commitments and model updates (cf. \cite{Agotnes07irrevocable} and \cite{Jamroga08commitment-tr}) and, more generally, requires involving strategy contexts in the semantics (\cite{BrihayeLLM09}).

\item Adding knowledge in the semantics, and explicitly in the language, by assuming that the agents reason and act under imperfect information.

\item Last, but most important long-term objective of this project is to model and capture by semantically richer logic-based formalism the \emph{mutually conditional strategic reasoning}, where all agents reason about their strategic choices, conditional on the others' strategic choices, conditional on the reasoners' choices, etc., recursively.
\end{enumerate}

\end{document}